\documentclass[aps,showpcs,amsmath,amssymb,longbibliography,twocolumn]{revtex4-1}
\usepackage{booktabs}      

\usepackage{array}
\usepackage{booktabs}
\usepackage{siunitx}
\usepackage[makeroom]{cancel}

\usepackage{multirow}
\usepackage{dashrule}

\usepackage{epsfig}
\usepackage{tikz}
\usetikzlibrary{quantikz}
\usepackage{amsmath}
\usepackage{bm,natbib}
\usepackage{times}
\usepackage{graphicx}
\usepackage{color}
\usepackage{slashed}
\usepackage{graphicx}
\usepackage{multirow}
\usepackage{amsmath,natbib} 
\usetikzlibrary{positioning,shapes}
\usepackage{relsize} 
\usepackage[utf8]{inputenc}
\usepackage{hyperref}  
\def\bea{\begin{eqnarray}}
\def\eea{\end{eqnarray}}
\def\bean{\begin{equation*}}
\def\eean{\end{equation*}} 

\begin{document}

\title{Production-Dependent Interpretation of the KLOE--MAMI $\eta\!\to\!\pi^{0}\gamma\gamma$ Tension\\[2pt] via the Nucleon-Triggered Leptophobic Vector $V_{\!\mathcal B}$}

\author{Yaroslav~Balytskyi}
\email{hr6998@wayne.edu, ybalytsk@uccs.edu}
\affiliation{Department of Physics and Astronomy, Wayne State University, Detroit, MI, 48201, USA}

\date{\today}

\begin{abstract}

The recent KLOE measurement $\mathrm{BR}^{\eta\rightarrow\pi^{0}\gamma\gamma}_{\text{KLOE}}  = (0.98\pm0.11_{\text{stat}}\pm0.14_{\text{syst}})\times10^{-4}$ is less than half the current world average, $(2.55\pm0.22)\times10^{-4}$, dominated by MAMI photoproduction data. We show that this $\approx 5.5\, \sigma$ discrepancy can be resolved by the new leptophobic, nucleon-triggered vector particle $V_{\mathcal B}$ with \(1.5~\text{GeV}\lesssim m_{V_{\!\mathcal B}}\lesssim5~\text{GeV}\), coupled via the effective operator  \(\bigl(\bar N N\bigr)\,\widetilde V_{\mathcal{B}}^{\mu\nu}F_{\mu\nu}P\).  This interaction modifies the \(\eta^{(\prime)}\!\to\!\pi^{0}\left(\eta\right)\gamma\gamma\) decay rates \textit{only} in the processes involving an external nucleon current, \(\gamma p\!\to\!\eta^{(\prime)}p\) and \(\pi^-p\!\to\!\eta^{(\prime)}n\), but leaves purely leptonic production channels, such as \(e^+e^-\to \phi\to\eta^{(\prime)}\gamma\) at KLOE and \(e^+e^-\to J/\psi\to\gamma\eta^{(\prime)}\) at BESIII, Standard-Model-like. The same mechanism predicts a \(\approx 10\%\) nucleon-triggered enhancement of the \(\eta^\prime\!\to\!\pi^{0}\gamma\gamma\) decay rate and a negligible shift for \(\eta^\prime\!\to\!\eta\gamma\gamma\), together with an \(\rm{A}^{2}\)-scaling boost if produced on heavy nuclei instead of protons. $V_{\!\mathcal B}$ can be searched directly in $2\!\to\!3$ photoproduction, for example, $\gamma p\!\to\!V_{\!\mathcal B}\,\pi^{0} p$. An integrated experimental program that compares $\eta^{(\prime)}\!\to\!\pi^{0}(\eta)\gamma\gamma$ in the presence of external nucleon currents with purely leptonic production, and conducts direct photoproduction searches for a GeV-scale vector, can decisively confirm or exclude our nucleon-rescaled, leptophobic–vector interpretation of the KLOE--MAMI discrepancy. 

\end{abstract}

\maketitle

\section{Introduction}

Precision measurements of long-lived neutral mesons $\eta$ and $\eta^{\prime}$ provide a clean laboratory for testing low-energy quantum chromodynamics (QCD) and, consequently, the search for New Physics (NP)~\cite{Gan:2020aco,nelson1989,fayet2006}. The doubly radiative decays \(\eta^{(\prime)} \rightarrow \pi^0(\eta)\, \gamma\gamma\) are particularly sharp probes of chiral perturbation theory (\(\chi\)PT) and its extensions. Since their decay spectra are dominated by intermediate vector mesons, $V\in\{\omega,\rho,\phi\}$, they are also highly sensitive to leptophobic New Physics scenarios~\cite{nelson1989,tulin2014}.

Landsberg~\cite{Landsberg:1985} summarized early data on the decay $\eta \rightarrow \pi^0 \gamma\gamma$. The GAMS-2000 experiment investigated this channel using the charge-exchange reaction $\pi^- + p \rightarrow \eta + n \rightarrow \left(\pi^0\gamma\gamma\right) + n$, reporting a branching ratio $\mathrm{BR}^{\eta\rightarrow\pi^0\gamma\gamma} = (7.1 \pm 1.4) \times 10^{-4}$~\cite{Alde1984}. Further studies with higher statistics, performed with Crystal Ball@AGS using the same reaction, measured $\mathrm{BR}^{\eta\rightarrow\pi^0\gamma\gamma} = (3.5 \pm 0.7 \pm 0.6) \times 10^{-4}$~\cite{Prakhov2005} in 2005, and in 2008, Prakhov \textit{et al.} reported $\mathrm{BR}^{\eta\rightarrow\pi^0\gamma\gamma} = (2.21 \pm 0.24 \pm 0.47) \times 10^{-4}$~\cite{Prakhov2008}, as well as the corresponding diphoton invariant mass spectrum. Independent reanalysis of the Crystal Ball data yielded a consistent value of $\mathrm{BR}^{\eta\rightarrow\pi^0\gamma\gamma} = (2.7 \pm 0.9 \pm 0.5) \times 10^{-4}$~\cite{Knecht2004}.

More recently, the A2 Collaboration at MAMI measured the decay via photoproduction, $\gamma + p \rightarrow \eta + p \rightarrow \left(\pi^0\gamma\gamma\right) + p$, reporting a partial decay width \(\Gamma(\eta \rightarrow \pi^0 \gamma\gamma) = (0.33 \pm 0.03)\,\mathrm{eV}\), based on a sample of approximately \(1.2 \times 10^3\) signal events~\cite{nefkens2014new}. Using the PDG value
for the total width, $\Gamma_\eta=1.31(5)\,\textrm{keV}$~\cite{PDG}, this corresponds to the branching ratio: 
\begin{equation}
\label{BRMAMI}
\mathrm{BR}^{\eta\to\pi^0\gamma\gamma}_{\text{MAMI}}=
\frac{\Gamma(\eta\to\pi^0\gamma\gamma)}{\Gamma_\eta}
=(2.52\pm0.25)\times10^{-4},
\end{equation}
where the uncertainty is obtained by standard error propagation:
\begin{equation}
\Delta\mathrm{BR}^{\eta\rightarrow\pi^0\gamma\gamma}
=\mathrm{BR}^{\eta\rightarrow\pi^0\gamma\gamma}
\sqrt{\left(\frac{\Delta\Gamma_\eta}
{\Gamma_\eta}\right)^2
+\left(\frac{\Delta\Gamma_{\eta\rightarrow\pi^0\gamma\gamma}}{\Gamma_{\eta\rightarrow\pi^0\gamma\gamma}}\right)^2}\,
\end{equation}

We note that, with the exception of the early GAMS-2000 measurement~\cite{Alde1984}, all measurements using initial nucleon states -- either charge-exchange reactions or photoproduction -- are \textit{internally consistent} and converge to a common range of approximately $\mathrm{BR}^{\eta\rightarrow\pi^0\gamma\gamma}\approx(2.2\ \text{-}\ 2.7) \times 10^{-4}$, leading to the current PDG world average~\cite{PDG}:
\begin{equation}
\label{BRPDG}
 \mathrm{BR}_{\text{PDG}}^{\eta \rightarrow \pi^0 \gamma\gamma} = (2.55 \pm 0.22) \times 10^{-4}
 \end{equation}

In sharp contrast, the KLOE Collaboration used the reaction \(e^+e^- \rightarrow \phi \rightarrow \eta \gamma \rightarrow (\pi^0 \gamma\gamma) + \gamma\), and has \emph{consistently} obtained a much lower branching ratio. Their initial result, \(\mathrm{BR}^{\eta \rightarrow \pi^0 \gamma\gamma} = (0.84 \pm 0.27 \pm 0.14) \times 10^{-4}\)~\cite{dimillo2006}, was confirmed by a recent 2025 update: \(\mathrm{BR}^{\eta \rightarrow \pi^0 \gamma\gamma}_{\mathrm{KLOE}} = (0.98 \pm 0.11_{\mathrm{stat}} \pm 0.14_{\mathrm{syst}}) \times 10^{-4}\)~\cite{Babusci2025}, and remains consistent with their earlier measurement~\cite{dimillo2006}. This value is approximately half of both the MAMI-derived branching ratio (Eqn.~\eqref{BRMAMI}) and the current PDG average (Eqn.~\eqref{BRPDG}), corresponding to \(5.0\,\sigma\) and \(5.5\,\sigma\) discrepancies, respectively. As the Authors note: ``This result agrees with a preliminary KLOE measurement, but is twice smaller than the present world average''~\cite{dimillo2006}.

A \textit{statistically significant} tension between the KLOE (leptonic) and MAMI (nucleon) measurements strongly suggests a \underline{\emph{production–dependent}} New Physics origin. Any viable New Physics explanation of this discrepancy must therefore couple specifically to external nucleons while leaving leptonic channels untouched. Further in the text, we develop such a \textit{nucleon-rescaled} scenario, in which the $\eta^{(\prime)}\!\to\!\pi^{0}(\eta)\gamma\gamma$ decay width is modified only when the nucleon current is present.

For the related decays, \(\eta^\prime \to \pi^0 \gamma\gamma\) and \(\eta^\prime \to \eta \gamma\gamma\), only the leptonic production channel has been explored so far, by BESIII, using the process \(e^+e^- \to J/\psi \to \gamma\eta^\prime\)~\cite{ablikim2017observation,ablikim2019search}. BESIII reports \(\mathrm{BR}_{\textrm{BESIII}}^{\eta^\prime \to \pi^0 \gamma\gamma} = \left(3.20 \pm 0.07 \pm 0.23\right) \times 10^{-3}\), along with the corresponding invariant diphoton mass spectrum~\cite{ablikim2017observation}. This value is higher than the upper bound \(\mathrm{BR}^{\eta^\prime \to \pi^0 \gamma\gamma} < 8 \times 10^{-4}\) at \(90\%\) confidence level (CL) previously reported by GAMS-2000~\cite{alde1987neutral}. For \(\eta^\prime \to \eta \gamma\gamma\), BESIII sets an upper limit of \(\mathrm{BR}_{\textrm{BESIII}}^{\eta^\prime \to \eta \gamma\gamma} < 1.33 \times 10^{-4}\) at \(90\%\) CL~\cite{ablikim2019search}.

Since our mechanism, discussed further in the text, changes the amplitude only in the presence of a nucleon current \((\bar{N}N)\), the BESIII results serve as clean Standard Model benchmarks, and are predicted to remain unchanged. In contrast, photoproduction and charge-exchange experiments that produce \(\eta^\prime\) in a nuclear environment should exhibit the same type of enhancement we propose for the \(\eta\), offering a fully falsifiable prediction that we quantify further in the text.

From a theoretical perspective, the computational landscape for the decay \(\eta \to \pi^0 \gamma\gamma\) encompasses a wide range of approaches. These include the Vector Meson Dominance (VMD) model~\cite{oppo1967models,baracca1970general}, chiral perturbation theory (\(\chi\)PT)~\cite{ametller1992chiral}, and its extensions incorporating \(C\)-odd axial-vector resonances~\cite{ko1993contributions,ko1995eta}. Other frameworks include unitarized chiral amplitudes~\cite{oset2003eta,oset2008eta}, dispersive methods~\cite{danilkin2017theoretical}, quark-loop (box) diagram calculations~\cite{ng1993,nemoto1996}, as well as both the original and extended versions of the Nambu--Jona-Lasinio model~\cite{belkov1995,bellucci1995,bijnens1995}. For the related decays, \(\eta^\prime \to \pi^0 \gamma\gamma\) and \(\eta^\prime \to \eta \gamma\gamma\), the preliminary theoretical predictions have been presented in~\cite{escribano2012,jora2010,Balytskyi:2018pzb,Balytskyi:2018uxb}, while a comprehensive treatment of all three channels simultaneously, \(\eta^{(\prime)} \to \pi^0(\eta) \gamma\gamma\), was performed in~\cite{escribano2020theoretical}. These decays were also analyzed using VMD in~\cite{Schaefer:2023stm}, yielding similar results to~\cite{escribano2020theoretical,Balytskyi:2023}. In this work, we use a modified approach from~\cite{escribano2020theoretical} as our Standard Model (SM) benchmark, which is consistent with the computation performed in our previous study~\cite{Balytskyi:2023}.

Our work is structured as follows.  Section~\ref{SMpart} reviews the Standard Model amplitude, which is the same in both leptonic and nucleon-induced production.  In Section~\ref{OurModel}, we present our production-dependent New Physics scenario, which becomes functional only in nucleon-induced production channels (photoproduction, charge-exchange). Section~\ref{Numerical} presents a fit to the MAMI $\eta\!\to\!\pi^{0}\gamma\gamma$ spectrum, extracting a single universal parameter that governs all three decays \(\eta^{(\prime)}\!\to\!\pi^{0}(\eta)\gamma\gamma\), up to a known group–theoretical factor. With this parameter fixed, we derive falsifiable predictions for \(\eta'\!\to\!\pi^{0}\gamma\gamma\) and \(\eta'\!\to\!\eta\gamma\gamma\) in nucleon production settings similar to MAMI or Crystal-Ball. In Section~\ref{Discussion}, we outline other falsifiable predictions of our model, and summarize our main conclusions in Section~\ref{Conclusions}.

\section{Standard Model Baseline}\label{SMpart}

First--principles predictions regarding the dynamics of $\eta$ and $\eta'$ mesons can be extracted from lattice QCD, and notable steps in this direction have already been reported~\cite{bali2021}. However, the current lattice-based calculations cannot yet reproduce all aspects of the low-energy behavior of these mesons. Therefore, we rely on phenomenological methods, the most important of which for the decays studied here is Vector Meson Dominance (VMD).

VMD originates from Sakurai's 1960 application of the Yang-Mills theory to strong interactions~\cite{sakurai1960}. Subsequently, Kroll, Lee, and Zumino incorporated electromagnetic VMD form factors in a gauge-invariant manner~\cite{kroll1960}. In this work, we follow the hidden local symmetry (HLS) implementation of VMD ~\cite{bando1985,bando19851,bando1988,fujiwara1985},
which provides an effective low-energy theory for the pseudoscalar meson
nonet $\bigl(\pi^{0},\eta,\eta',K,\bar K\bigr)$
together with the vector meson nonet
$\bigl(\rho^{0},\omega,\phi,K^{\!*},\bar K^{\!*}\bigr)$, treating the latter as gauge bosons of the hidden symmetry $\mathrm{U}(3)_{V}$. In this framework, hadronic amplitudes originate from a single vector--pseudoscalar--vector vertex whose coupling constant is fixed by the anomaly. The mixing of $V - \gamma$ is proportional to $e \rm{Tr}\left[\bf{Q}\cdot\bf{T_V}\right]$, where $\bf{Q}=diag\left(\frac{2}{3},-\frac{1}{3},-\frac{1}{3}\right)$ is the quark charge matrix, and $\bf{T_V}$ is the $\rm{U}\left(3\right)$ generator for $V$.

A consistent description at low energies also requires the incorporation of HLS into chiral perturbation theory ($\chi$PT), pioneered by Weinberg~\cite{weinberg1979} and Gasser and  Leutwyler~\cite{gasser1984,gasser1985}. Its extension that includes explicit resonant fields, resonant chiral theory (R$\chi$T)~\cite{RChTReview}, may also shed new light on the phenomenology of $\eta$/$\eta^\prime$.

The decay $\eta\!\to\!\pi^{0}\gamma\gamma$ is a rigorous test of the predictive ability of $\chi$PT~\cite{ametller1992chiral}. Since all participating pseudoscalars are neutral, the tree-level contributions at $O(p^{2})$ and $O(p^{4})$ are zero. At $O(p^{4})$, only kaon and pion loops contribute, but the latter are strongly suppressed by $G$--parity breaking $\propto m_{u}-m_{d}$. The only sizable effect comes at $O(p^{6})$, and the associated low-energy constants are fixed by matching to VMD, expanding the vector propagators in powers of $t/M_{V}^{2}$ and $u/M_{V}^{2}$~\cite{ametller1992chiral}, while the
$O(p^{8})$ two-anomaly loop corrections are negligible.

Extending these considerations to all three doubly radiative channels, $\eta^{\left(\prime\right)}\rightarrow\pi^0\left(\eta\right)\gamma\gamma$, our analysis follows the strategy of~\cite{escribano2020theoretical}: we work in the large-$N_{c}$ and isospin limits and consider the singlet $\eta_{0}$ as the ninth Goldstone pseudo-boson. By ``isospin limit'' we mean $m_{u}=m_{d}$ in the hadronic sector, so that the terms $\propto (m_{u}-m_{d})$ and isospin-violating mass splittings in hadronic thresholds are neglected. Since the decays are electromagnetic, exact conservation of isospin is \textit{not} expected, and so we do \textit{not} impose isospin as a symmetry constraint on the full amplitude. The vector-pseudoscalar-photon (VP$\gamma$) couplings are implemented in the standard VMD parametrization, with on-shell $g_{V P \gamma}$ extracted from $V\!\to\!P\gamma$ or $P\!\to\!V\gamma$ decay widths. 

Under these assumptions, only kaon loops contribute to $\eta^{\prime}\!\to\!\pi^{0}\gamma\gamma$ and
$\eta\!\to\!\pi^{0}\gamma\gamma$, while both kaon and pion loops are present in $\eta^{\prime}\!\to\!\eta\gamma\gamma$. Vector mesons $V = \{\omega,\rho^0,\phi\}$ entering through the cascade \(\eta^{(\prime)} \to V\gamma \to \pi^{0} \gamma \gamma\) are modeled within the framework of VMD using the same $g_{V P \gamma}$ couplings, while scalar resonances are explicitly included through the Linear Sigma Model ($\mathrm{L}\sigma\mathrm{M}$). 

The complementarity between $\mathrm{L}\sigma\mathrm{M}$ and $\chi$PT preserves the correct low-energy behavior -- a recipe that has already been confirmed in the $V\!\to\!P^{0}P^{0}\gamma$ decays~\cite{escribano2006}. Finally, loop graphs containing two anomalous vertices give a negligible contribution and are therefore omitted. The overall SM matrix element is a coherent sum of VMD+L$\sigma$M contributions:

\begin{equation}
    \lvert \mathcal{M}\lvert^2 =   \lvert \mathcal{M}^{\rm{VMD}}\lvert^2   +  2\rm{Re} \left(\mathcal{M}^{\rm{VMD}}\left(\mathcal{M}^{L \sigma M}\right)^\dagger\right)+\lvert \mathcal{M}^{L \sigma M}\lvert^2
\end{equation}
Similarly to~\cite{escribano2020theoretical,Balytskyi:2023}, our calculation assumes no relative phase between the VMD and L\(\sigma\)M contributions. In any case, the introduction of such a phase only leads to a difference in the predicted decay width $\eta\to\pi^0\gamma\gamma$ of only a few percent~\cite{Balytskyi:2023}, which is not enough to explain more than twofold discrepancy between the KLOE and MAMI measurements~\cite{nefkens2014new,Babusci2025}.

The Lorentz-invariant structures $\{a\}$ and $\{b\}$ defining the matrix element are:
\begin{equation}
\begin{aligned}
&\{a\} = (\epsilon_1\cdot\epsilon_2)(q_1\cdot q_2)-(\epsilon_1\cdot q_2)(\epsilon_2\cdot q_1) \ , \\
&\{b\} =(\epsilon_1\cdot q_2)(\epsilon_2\cdot P)(P\cdot q_1)+(\epsilon_2\cdot q_1)(\epsilon_1\cdot P)(P\cdot q_2) -\\
&-(\epsilon_1\cdot\epsilon_2)(P\cdot q_1)(P\cdot q_2)-(\epsilon_1\cdot P)(\epsilon_2\cdot P)(q_1\cdot q_2), \ 
\end{aligned}
\end{equation}
where \(P\) is the four-momentum of the decaying $\eta^{\left(\prime\right)}$, and the final-state particles are labeled as \(\{1,2,3\} = \{\gamma, \gamma, \pi^0(\eta)\}\), and \(\epsilon_{1,2}\) and \(q_{1,2}\) are the polarization vectors and four-momenta of the two outcoming photons, respectively.

\begin{figure}[h!] 
\includegraphics[width=0.5\textwidth]{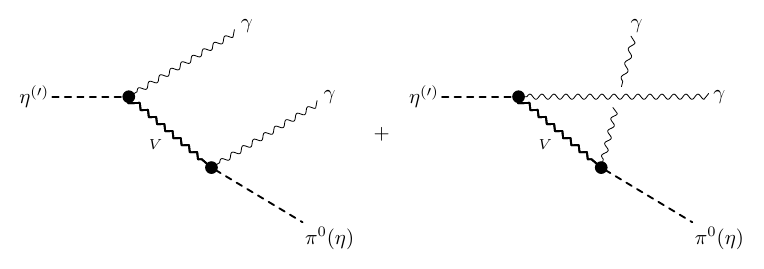}
\caption{The dominant VMD diagrams contributing to the $\eta^{\left(\prime\right)}\rightarrow\pi^0\left(\eta\right)\gamma\gamma$ decays, corresponding to Eqn.~\eqref{VMD}.}
\label{Diagram}
\end{figure}

The dominant VMD contribution governing all three decays $\eta^{\left(\prime\right)}\rightarrow\pi^0\left(\eta\right)\gamma\gamma$ is shown in Eqn.~\eqref{VMD} and Fig.~(\ref{Diagram}):

\begin{widetext}
\begin{eqnarray}\label{VMD}
\quad {\cal M}^{\mathrm{VMD}}_{\eta^{\left(\prime\right)}\to\pi^0\left(\eta\right)\gamma\gamma}=
\sum_{V=\rho^0, \omega, \phi}g_{V\!\eta^{\left(\prime\right)}\gamma}g_{V\!\pi^0\left(\eta\right)\gamma}\left[\frac{(P\cdot q_2-m_{\eta^{\left(\prime\right)}}^2)\{a\}-\{b\}}{D_V(t)}+
\bigg\{
\begin{array}{c}
q_2\leftrightarrow q_1\\
t\leftrightarrow u
\end{array}
\bigg\}\right]\ ,
\end{eqnarray}
\end{widetext}
where $t,u=(P-q_{2,1})^2=m^2_{\eta^{\left(\prime\right)}}-2P\cdot q_{2,1}$ are the Mandelstam variables.

Note that, in general, the sum of Breit-Wigner functions in Eqn.~\eqref{VMD} leads to a violation of unitarity constraints~\cite{PDG-Resonances-2024}. However, for the case of $\eta\rightarrow\pi^0\gamma\gamma$ and $\eta^\prime\rightarrow\eta\gamma\gamma$, $\omega$ and $\rho^0$ resonances, as well as the New Physics contribution, considered further in the text, are all far off-shell, and therefore the unitarity-violating effects of the Breit-Wigner sum are expected to be numerically insignificant at our level of accuracy.

In contrast, $\omega$ and $\rho^0$ can go on shell in
$\eta^\prime\rightarrow\pi^0\gamma\gamma$. This decay is nevertheless dominated by a single narrow $\omega$ peak, sitting on top of a slowly varying $\rho^0$ contribution. Therefore, we expect the unitarity violations to be moderate. We defer a more sophisticated, manifestly unitary approach to our future work.

In the exact $\mathrm{SU}(3)$ flavor symmetry and OZI limits, 
all the $g_{VP\gamma}$ couplings can be reduced to a single overall constant multiplied by group-theoretical factors~\cite{bramon1995}. To accommodate the residual effects, we adopt the phenomenological framework developed for $V\!\to\!P\gamma$ and $P\!\to\!V\gamma$
decays~\cite{Bramon:2000fr,Escribano:2020jdy}. In this approach, the mismatch between the effective magnetic moments of the light $u/d$ and strange $s$ constituent quarks is encoded through their mass ratio, and the $s$-quark entry in the charge matrix ${\bf Q}$ is multiplied by the factor $1 - s_e \equiv \frac{\bar{m}}{m_s}$. This single parameter $s_{e}$ quantifies the amount of $\mathrm{SU}(3)$–flavor breaking and, simultaneously, the deviation from ideal OZI in the $VP\gamma$ vertices.

The resulting coupling constants $g_{VP\gamma}$ that we use are as follows:
\begin{equation}
\label{CouplingConstants}
\begin{aligned}
& g_{\rho^0\pi^0\gamma} = \frac{g}{3}, \ g_{\rho^0\eta\gamma} = gz_{\textrm{NS}}\cos{\varphi_P}, \ g_{\rho^0\eta^{\prime}\gamma} = gz_{\textrm{NS}}\sin{\varphi_P},
\\ 
& g_{\omega \pi^0 \gamma} = g\cos{\varphi_V}, \\
& g_{\omega\eta\gamma} = \frac{g}{3}\left(z_{\textrm{NS}}\cos{\varphi_P}\cos{\varphi_V} - 2 \frac{\overline{m}}{m_s}z_{\rm{S}}\sin{\varphi_P}\sin{\varphi_V}\right), \\
& g_{\omega\eta^{\prime}\gamma} = \frac{g}{3}\left(z_{\textrm{NS}}\sin{\varphi_P}\cos{\varphi_V} + 2 \frac{\overline{m}}{m_s}z_{\rm{S}}\cos{\varphi_P}\sin{\varphi_V}\right), \\
& g_{\phi\pi^0\gamma} = g\sin{\varphi_V}, \\
& g_{\phi\eta\gamma} = \frac{g}{3}\left(z_{\rm{NS}}\cos{\varphi_P}\sin{\varphi_V} + 2 \frac{\overline{m}}{m_s}z_{\rm{S}}\sin{\varphi_P}\cos{\varphi_V}\right), \\
& g_{\phi\eta^{\prime}\gamma} = \frac{g}{3}\left(z_{\rm{NS}}\sin{\varphi_P}\sin{\varphi_V} - 2 \frac{\overline{m}}{m_s}z_{\rm{S}}\cos{\varphi_P}\cos{\varphi_V}\right)\ 
\end{aligned}
\end{equation}
Here, \(g\) denotes a generic electromagnetic coupling constant, and \(\varphi_P\) is the pseudoscalar \(\eta\text{--}\eta^\prime\) mixing angle in the quark-flavor basis, defined at the leading order in \(\chi\)PT as:
\begin{equation}
\begin{cases}
\ket{\eta} = \ket{\eta_{NS}}\cos\left(\varphi_P\right) - \ket{\eta_S}\sin\left(\varphi_P\right) \\
\ket{\eta^\prime} = \ket{\eta_{NS}}\sin\left(\varphi_P\right) + \ket{\eta_S}\cos\left(\varphi_P\right) 
\end{cases},
\end{equation}
with $\ket{\eta_{NS}} = \frac{1}{\sqrt{2}}\left(\ket{u\bar{u}} + \ket{d\bar{d}}\right)$ and $\ket{\eta_S} = \ket{s\bar{s}}$~\cite{bramon1997}. $\varphi_V$ denotes the analogous \(\omega\text{--}\phi\) mixing angle in the same basis. The differences in the spatial wavefunction overlaps of the light and strange quark are encoded by the dimensionless factors \(z_{NS}\) and \(z_{S}\), respectively. The numerical values of the above-mentioned coefficients are as follows~\cite{Escribano:2020jdy}:

\begin{equation}\label{fit4}
\begin{aligned}
& g = 0.70(1) \ \rm{GeV}^{-1},	\varphi_{P} = 41.4(5)^\circ, z_{\rm{NS}} = 0.83(2), \\
& z_{\rm{S}}\overline{m}/m_s = 0.65(1),  \varphi_{V} = 3.3(1)^\circ
\end{aligned}
\end{equation}
For completeness, Table~\rm{I} summarizes the theoretical predictions and the corresponding experimental results for the $V\rightarrow P\gamma$ and $P\rightarrow V\gamma$ decays, using the values in Eqn.~\eqref{fit4}. These values have already been employed in~\cite{escribano2020theoretical,Balytskyi:2023}.

\begin{table}
\begin{center}
\begin{tabular}{|c|c|c|} 
\hline
Decay  & $\Gamma^{exp.} \left(\rm{keV}\right)$ & $\Gamma^{th.} \left(\rm{keV}\right)$ \\
\hline
$\rho^0\rightarrow\eta\gamma$ &  $44\left(3\right)$ & $38\left(2\right)$ \\ \hline
$\rho^0\rightarrow\pi^0\gamma$ &  $69\left(9\right)$ & $79\left(2\right)$  \\ 
\hline
$\omega\rightarrow\eta\gamma$ & $3.8\left(3\right)$ & $3.5\left(2\right)$   \\
\hline
$\omega\rightarrow\pi^0\gamma$ & $713\left(20\right)$ & $704\left(19\right)$   \\
\hline
$\phi\rightarrow\eta\gamma$ & $54.4\left(1.1\right)$ & $54\left(8\right)$   \\
\hline
$\phi\rightarrow\eta^\prime\gamma$ & $0.26\left(1\right)$ & $0.27\left(5\right)$   \\
\hline
$\phi\rightarrow\pi^0\gamma$ & $5.5 \left(2\right)$ & $ 5.5\left(3\right)$   \\
\hline
$\eta^\prime\rightarrow\rho^0\gamma$ & $57\left(3\right)$ & $55\left(3\right)$   \\
\hline
$\eta^\prime\rightarrow\omega\gamma$ & $5.1 \left(3\right)$ & $6.5\left(1\right)$   \\ \hline
\end{tabular}
\caption{Summary on the $V\rightarrow P\gamma$ and $P\rightarrow V\gamma$ decays using the phenomenological model from~\cite{Bramon:2000fr,Escribano:2020jdy}, with numerical inputs taken from the fit values in Eqn.~\eqref{fit4}.}
\end{center}
\end{table}

In our computations, we keep the subleading L$\sigma$M contribution to the Standard Model amplitude exactly as given in~\cite{escribano2020theoretical}. For completeness, this contribution is summarized below in its original form.

\begin{widetext}

\begin{equation}
\label{AKpKmpi0etaChPTLsM}
\begin{aligned}
{\cal A}_{K^+K^- \to \pi^0\eta}^{\mbox{\scriptsize L$\sigma$M}} =
\frac{1}{2f_\pi f_K} \bigg\{ 
&(s - m_\eta^2) \frac{m_K^2 - m_{a_0}^2}{D_{a_0}(s)} \cos\varphi_P 
+ \frac{1}{6} \Big[ 
(5m_\eta^2 + m_\pi^2 - 3s) \cos\varphi_P \\
&\quad - \sqrt{2}(m_\eta^2 + 4m_K^2 + m_\pi^2 - 3s) \sin\varphi_P 
\Big] \bigg\} \,,
\end{aligned}
\end{equation}

\begin{equation}
\label{Apipimpi0etapChPTLsM}
\begin{aligned}
{\cal A}_{K^+K^- \to \pi^0\eta'}^{\mbox{\scriptsize L$\sigma$M}} =
\frac{1}{2f_\pi f_K} \bigg\{ 
&(s - m_{\eta'}^2) \frac{m_K^2 - m_{a_0}^2}{D_{a_0}(s)} \sin\varphi_P 
+ \frac{1}{6} \Big[ 
(5m_{\eta'}^2 + m_\pi^2 - 3s) \sin\varphi_P \\
&\quad + \sqrt{2}(m_{\eta'}^2 + 4m_K^2 + m_\pi^2 - 3s) \cos\varphi_P 
\Big] \bigg\} \,,
\end{aligned}
\end{equation}

\begin{equation}
\label{AKpKmetaetapChPTLsM}
\begin{aligned}
{\cal A}_{K^+K^- \to \eta\eta'}^{\mbox{\scriptsize L$\sigma$M}} =
&\frac{s - m_K^2}{2f_K} \bigg[
\frac{g_{\sigma\eta\eta'}}{D_\sigma(s)} \left( \cos\varphi_S - \sqrt{2} \sin\varphi_S \right)
+ \frac{g_{f_0\eta\eta'}}{D_{f_0}(s)} \left( \sin\varphi_S + \sqrt{2} \cos\varphi_S \right)
\bigg] \\[1ex]
&- \frac{s - m_K^2}{4f_\pi f_K}
\left[ 1 - 2\left( \frac{2f_K}{f_\pi} - 1 \right) \right] \sin(2\varphi_P) \\[1ex]
&- \frac{1}{4f_\pi^2} \bigg[
\left( s - \frac{m_\eta^2 + m_{\eta'}^2}{3} - \frac{8m_K^2}{9} - \frac{2m_\pi^2}{9} \right)
\left( \sqrt{2} \cos(2\varphi_P) + \frac{1}{2} \sin(2\varphi_P) \right) \\
&\quad + \frac{4}{9}(2m_K^2 - m_\pi^2)
\left( 2\sin(2\varphi_P) - \frac{1}{\sqrt{2}} \cos(2\varphi_P) \right)
\bigg] \,,
\end{aligned}
\end{equation}

\begin{equation}
\label{ApippimetaetapChPTLsM}
\begin{aligned}
{\cal A}_{\pi^+\pi^- \to \eta\eta'}^{\mbox{\scriptsize L$\sigma$M}} =
\frac{s - m_\pi^2}{f_\pi} \bigg[
\frac{g_{\sigma\eta\eta'}}{D_\sigma(s)} \cos\varphi_S
+ \frac{g_{f_0\eta\eta'}}{D_{f_0}(s)} \sin\varphi_S
\bigg]
+ \frac{2m_\pi^2 - s}{2f_\pi^2} \sin(2\varphi_P) \,,
\end{aligned}
\end{equation}

\vspace{1.5ex}

\begin{equation}
\label{gsigmaetaetap}
\begin{aligned}
g_{\sigma\eta\eta'} = \frac{\sin(2\varphi_P)}{2f_\pi} \bigg\{
&(m_\eta^2 \cos^2\varphi_P + m_{\eta'}^2 \sin^2\varphi_P - m_{a_0}^2)
\left[ \cos\varphi_S + \sqrt{2} \sin\varphi_S \left( 2\frac{f_K}{f_\pi} - 1 \right) \right] \\[1ex]
&- (m_{\eta'}^2 - m_\eta^2)
\left( \cos\varphi_S \cos(2\varphi_P) - \frac{1}{2} \sin\varphi_S \sin(2\varphi_P) \right)
\bigg\} \,,
\end{aligned}
\end{equation}

\begin{equation}
\label{gf0etaetap}
\begin{aligned}
g_{f_0\eta\eta'} = \frac{\sin(2\varphi_P)}{2f_\pi} \bigg\{
&(m_\eta^2 \cos^2\varphi_P + m_{\eta'}^2 \sin^2\varphi_P - m_{a_0}^2)
\left[ \sin\varphi_S - \sqrt{2} \cos\varphi_S \left( 2\frac{f_K}{f_\pi} - 1 \right) \right] \\[1ex]
&- (m_{\eta'}^2 - m_\eta^2)
\left( \sin\varphi_S \cos(2\varphi_P) + \frac{1}{2} \cos\varphi_S \sin(2\varphi_P) \right)
\bigg\} \,.
\end{aligned}
\end{equation}

\end{widetext}

The four-pseudoscalar amplitudes and the corresponding couplings are presented in Eqns.~\eqref{AKpKmpi0etaChPTLsM},~\eqref{Apipimpi0etapChPTLsM},~\eqref{AKpKmetaetapChPTLsM},~\eqref{ApippimetaetapChPTLsM},~\eqref{gsigmaetaetap}, and Eqn.~\eqref{gf0etaetap}. 

The scalar mixing angle is numerically fixed as $\varphi_S=-8^\circ$, following~\cite{escribano2006,escribano2020theoretical}, and is defined as: 
\begin{equation}
\begin{aligned}
\begin{cases}
\ket{\sigma} = 
\phantom{-} \cos\varphi_{S}\, \ket{\sigma_{\!NS}} 
- \sin\varphi_{S}\, \ket{\sigma_{\!S}} \\[4pt]
\ket{f_0} = 
\phantom{-} \sin\varphi_{S}\, \ket{\sigma_{\!NS}} 
+ \cos\varphi_{S}\, \ket{\sigma_{\!S}}
\end{cases},
\end{aligned}
\end{equation}
where the non-strange and strange flavor components are defined as \(\ket{\sigma_{\!NS}}=\tfrac{1}{\sqrt{2}}\bigl(\ket{u\bar u}+\ket{d\bar d}\bigr)\) and
\(\ket{\sigma_{\!S}}=\ket{s\bar s}\).

The L$\sigma$M contribution to the $\eta^{\left(\prime\right)}\rightarrow\pi^0\gamma\gamma$ decays contains only the kaon loop:
\begin{equation}
{\cal M}^{\mathrm{L\sigma M}}_{\eta^{\left(\prime\right)}\to\pi^0\gamma\gamma} =
\frac{2\alpha}{\pi}\frac{1}{m_{K^+}^2}L(s_K)\{a\}\times{\cal M}^{\rm{L\sigma M}}_{K^+K^-\to\pi^0\eta^{\left(\prime\right)}}
\end{equation}
Both pion and kaon loops contribute to the decay
\(\eta^{\prime}\!\to\!\eta\gamma\gamma\): 
\begin{multline}
{\cal M}^{\mathrm{L\sigma M}}_{\eta^\prime\to\eta\gamma\gamma} = 
\frac{2\alpha}{\pi}\frac{1}{m_{\pi}^2}L(s_\pi)\{a\}
\times{\cal M}^{\rm{L\sigma M}}_{\pi^+\pi^-\to\eta\eta^{\prime}} + \\
+ \frac{2\alpha}{\pi}\frac{1}{m_{K^+}^2}L(s_K)\{a\}
\times{\cal M}^{\rm{L\sigma M}}_{K^+K^-\to\eta\eta^{\prime}}
\end{multline}
The corresponding loop integral is:
\begin{equation}
\begin{aligned}
&L\left(z\right)=-\frac{1}{2z}-\frac{2}{z^2}f\left(\frac{1}{z}\right),\\
&f\left(z\right)=
\begin{cases}{}
\frac{1}{4}\left(\log\frac{1+\sqrt{1-4z}}{1-\sqrt{1-4z}}-i\pi\right)^2, & \mbox{if}\ z<\frac{1}{4}\\[1ex]
-\left[\arcsin\left(\frac{1}{2\sqrt{z}}\right)\right]^2, & \mbox{if}\ z>\frac{1}{4}
\end{cases},
\end{aligned}
\end{equation}
where $s_K=s/m_{K^+}^2$, $s_\pi=s/m_{\pi^+}^2$, and $s=(q_1+q_2)^2=2q_1\cdot q_2$ represents the invariant mass of the two outgoing photons.

$a_{0}(980)$ contributes to both $\eta^{(\prime)}\!\to\!\pi^{0}\gamma\gamma$ decays, and its renormalized mass is fixed as $m_{a_{0}} = 980~\rm{MeV}$. For the decay $\eta^{\prime}\!\to\!\eta\gamma\gamma$, the contributing scalar states are $\sigma(500)$ and $f_{0}(990)$, with $m_{\sigma}=498~\rm{MeV}$ and $m_{f_{0}}=990~\rm{MeV}$. We employ the physical decay constants as
$f_{\pi}=92.07~\text{MeV}$ and $f_{K}=110.10~\text{MeV}$. Each scalar resonance is represented by the dressed propagator as:
\begin{equation}
\label{ScalarResonances}
  D_{R}(s)=
  s-m_{R}^{2}
  +\text{Re}\left(\Pi_{R}(s)\right)-\text{Re}\left(\Pi_{R}(m_{R}^{2})\right)
  +i\,\text{Im}\left(\Pi_{R}(s)\right),
\end{equation}
with $\text{Re}\left(\Pi_{R}(s)\right)$ and $\text{Im}\left(\Pi_{R}(s)\right)$ defined below.

For $a_0$, the parameters are fixed as follows. The couplings of $a_0$ to kaons in the isospin limit are~\cite{escribano2020theoretical}:

\begin{equation}
\begin{cases}
g_{a_0 K\bar K}^2=2g_{a_0 K^+K^-}^2=\frac{1}{2}\left(\frac{m_{K}^2-m_{a_0}^2}{f_{K}}\right)^2 \\
g_{a_0 \pi \eta}^2=\left(\frac{m_{\eta}^2-m_{a_0}^2}{f_{\pi}}\cos{\varphi_P}\right)^2 \\
g_{a_0 \pi \eta^\prime}^2=\left(\frac{m_{\eta^\prime}^2-m_{a_0}^2}{f_{\pi}}\sin{\varphi_P}\right)^2
\end{cases}.
\end{equation}

The kinematic coefficients are defined as follows: $\beta_K=\sqrt{1-4m_K^2/s}$,
$\bar\beta_K=\sqrt{4m_K^2/s-1}$, $\theta_K=\theta(s-4m_K^2)$, and 
$\bar\theta_K=\theta(4m_K^2-s)$, $\beta^\pm_{\pi\eta^{\left(\prime\right)}}=\sqrt{1-(m_\pi\pm m_{\eta^{\left(\prime\right)}})^2/s}$,
$\bar\beta^\pm_{\pi\eta^{\left(\prime\right)}}=\sqrt{(m_\pi\pm m_{\eta^{\left(\prime\right)}})^2/s-1}$,
$\theta_{\pi\eta^{\left(\prime\right)}}=\theta[s-(m_\pi+m_{\eta^{\left(\prime\right)}})^2]$,
$\bar\theta_{\pi\eta^{\left(\prime\right)}}=\theta[s-(m_\pi-m_{\eta^{\left(\prime\right)}})^2]\times\theta[(m_\pi+m_{\eta^{\left(\prime\right)}})^2-s]$, and  $\bar{\bar\theta}_{\pi\eta^{\left(\prime\right)}}=\theta[(m_\pi-m_{\eta^{\left(\prime\right)}})^2-s]$.

The real and imaginary parts of the $a_0$ propagator are provided in Eqns.~\eqref{pi0Rea0} and~\eqref{pi0Ima0}.

\begin{widetext}

\begin{equation}
\label{pi0Rea0}
\begin{aligned}
\mathrm{Re}_{a_0}^{\eta^{\left(\prime\right)}\rightarrow\pi^0\gamma\gamma}\left(\Pi(s)\right) = 
&\;\frac{g_{a_0 K\bar{K}}^2}{16\pi^2}
\left[
2 - \beta_K \log\left(\frac{1 + \beta_K}{1 - \beta_K}\right) \theta_K
- 2\bar\beta_K \arctan\left(\frac{1}{\bar\beta_K}\right) \bar\theta_K
\right] \\
&+ \frac{g_{a_0 \pi\eta^{(\prime)}}^2}{16\pi^2}
\Bigg[
2 - \frac{m^2_{\eta^{(\prime)}} - m^2_\pi}{s}
     \log\left(\frac{m_{\eta^{(\prime)}}}{m_\pi}\right)
- \beta^+_{\pi\eta^{(\prime)}} \beta^-_{\pi\eta^{(\prime)}}
     \log\left(\frac{\beta^-_{\pi\eta^{(\prime)}} + \beta^+_{\pi\eta^{(\prime)}}}
                    {\beta^-_{\pi\eta^{(\prime)}} - \beta^+_{\pi\eta^{(\prime)}}}\right)
     \theta_{\pi\eta^{(\prime)}} \\
&\quad - 2\bar\beta^+_{\pi\eta^{(\prime)}} \beta^-_{\pi\eta^{(\prime)}}
       \arctan\left(\frac{\beta^-_{\pi\eta^{(\prime)}}}{\bar\beta^+_{\pi\eta^{(\prime)}}}\right)
       \bar\theta_{\pi\eta^{(\prime)}}
+ \bar\beta^+_{\pi\eta^{(\prime)}} \bar\beta^-_{\pi\eta^{(\prime)}}
     \log\left(\frac{\bar\beta^+_{\pi\eta^{(\prime)}} + \bar\beta^-_{\pi\eta^{(\prime)}}}
                    {\bar\beta^+_{\pi\eta^{(\prime)}} - \bar\beta^-_{\pi\eta^{(\prime)}}}\right)
     \bar{\bar\theta}_{\pi\eta^{(\prime)}}
\Bigg] \,,
\end{aligned}
\end{equation}

\vspace{-2ex}

\begin{equation}
\label{pi0Ima0}
\mathrm{Im}_{a_0}^{\eta^{\left(\prime\right)}\rightarrow\pi^0\gamma\gamma}\left(\Pi(s)\right) =
-\frac{g_{a_0 K\bar{K}}^2}{16\pi} \beta_K \theta_K
-\frac{g_{a_0 \pi\eta}^2}{16\pi} \beta^+_{\pi\eta^{\left(\prime\right)}} \beta^-_{\pi\eta} \theta_{\pi\eta} \,.
\end{equation}
\end{widetext}

\begin{widetext}
For $\sigma\left(500\right)$, the corresponding coupling constants are as follows:
\begin{equation}
\begin{cases}
g_{\sigma\pi\pi}^2 = \dfrac{3}{2}\,g_{\sigma\pi^+\pi^-}^2 
= \dfrac{3}{2} \left( \dfrac{m_\pi^2 - m_\sigma^2}{f_\pi} \cos\varphi_S \right)^2 \\
g_{\sigma K\bar{K}}^2 = 2\,g_{\sigma K^+K^-}^2 
= \dfrac{1}{2} \left[ \dfrac{m_K^2 - m_\sigma^2}{f_K} 
\left( \cos\varphi_S - \sqrt{2} \sin\varphi_S \right) \right]^2
\end{cases}
\end{equation}

And, finally, for $f_0\left(990\right)$: 

\begin{equation}
\begin{cases}
g_{f_0\pi\pi}^2 = \dfrac{3}{2}\,g_{f_0\pi^+\pi^-}^2 
= \dfrac{3}{2} \left( \dfrac{m_\pi^2 - m_{f_0}^2}{f_\pi} \sin\varphi_S \right)^2 \\  
g_{f_0 K\bar{K}}^2 = 2\,g_{f_0 K^+K^-}^2 
= \dfrac{1}{2} \left[ \dfrac{m_K^2 - m_{f_0}^2}{f_K} 
\left( \sin\varphi_S + \sqrt{2} \cos\varphi_S \right) \right]^2
\end{cases}
\end{equation}

The corresponding real and imaginary parts of the $\sigma\left(500\right)$ and $f_0\left(990\right)$ propagators have the following form: 

\begin{equation}
\label{EtaEtaPrimeSigmaR}
\begin{aligned}
R^{\eta^\prime\rightarrow\eta\gamma\gamma}_\sigma(s) = 
&\;\frac{g_{\sigma\pi\pi}^2}{16\pi^2}
\bigg[
2 - \beta_\pi \log\left( \frac{1 + \beta_\pi}{1 - \beta_\pi} \right) \theta_\pi
- 2\bar\beta_\pi \arctan\left( \frac{1}{\bar\beta_\pi} \right) \bar\theta_\pi
\bigg] \\[2pt]
&+ \frac{g_{\sigma K\bar{K}}^2}{16\pi^2}
\bigg[
2 - \beta_K \log\left( \frac{1 + \beta_K}{1 - \beta_K} \right) \theta_K
- 2\bar\beta_K \arctan\left( \frac{1}{\bar\beta_K} \right) \bar\theta_K
\bigg] \,,
\end{aligned}
\end{equation}

\vspace{-2ex}

\begin{equation}
\label{EtaEtaPrimeSigmaI}
I^{\eta^\prime\rightarrow\eta\gamma\gamma}_\sigma(s) = 
- \frac{g_{\sigma\pi\pi}^2}{16\pi} \beta_\pi \theta_\pi
- \frac{g_{\sigma K\bar{K}}^2}{16\pi} \beta_K \theta_K \,,
\end{equation}

\begin{equation}
\label{EtaEtaPrimeRf0}
\begin{aligned}
R^{\eta^\prime\rightarrow\eta\gamma\gamma}_{f_0}(s) = 
&\;\frac{g_{f_0\pi\pi}^2}{16\pi^2}
\bigg[
2 - \beta_\pi \log\left( \frac{1 + \beta_\pi}{1 - \beta_\pi} \right) \theta_\pi
- 2\bar\beta_\pi \arctan\left( \frac{1}{\bar\beta_\pi} \right) \bar\theta_\pi
\bigg] \\[2pt]
&+ \frac{g_{f_0 K\bar{K}}^2}{16\pi^2}
\bigg[
2 - \beta_K \log\left( \frac{1 + \beta_K}{1 - \beta_K} \right) \theta_K
- 2\bar\beta_K \arctan\left( \frac{1}{\bar\beta_K} \right) \bar\theta_K
\bigg] \,,
\end{aligned}
\end{equation}

\vspace{-2ex}

\begin{equation}
\label{EtaEtaPrimeIf0}
I^{\eta^\prime\rightarrow\eta\gamma\gamma}_{f_0}(s) =
- \frac{g_{f_0\pi\pi}^2}{16\pi} \beta_\pi \theta_\pi
- \frac{g_{f_0 K\bar{K}}^2}{16\pi} \beta_K \theta_K \,,
\end{equation}

\end{widetext}

It is worth noting that Eqn.~\eqref{ScalarResonances} underestimates the width of the very broad $f_{0}(500)$  (also called $\sigma\left(500\right)$), yielding at most $\Gamma_\sigma\!\approx\!132\,\text{MeV}$ at $s=(m_{\eta'}-m_\eta)^2$, which is a factor of $\sim 2\,-\,2.5$ less than the experimental width of the pole~\cite{PDG}. This has no impact on $\eta\!\to\!\pi^0\gamma\gamma$ and $\eta^\prime\!\to\!\pi^0\gamma\gamma$, where the $\sigma$ does not contribute, but may induce a few-percent bias in $\eta^\prime\!\to\!\eta\gamma\gamma$, where scalar exchange is subleading yet non-negligible.

The main conclusions of our paper do not depend on this, since we focus on the $\eta\!\to\!\pi^0\gamma\gamma$ decay, while $\eta^\prime\!\to\!\eta\gamma\gamma$ is only weakly sensitive to the New Physics contribution we consider further in the text. We acknowledge this limitation of the SM benchmark we use and defer a more refined treatment of the $\sigma\left(500\right)$ contribution to our future work.

In the next Section~\ref{OurModel}, we supplement these expressions with our proposed nucleon-triggered leptophobic dark photon model.

\section{Overview of our proposed model}
\label{OurModel}
To explain the KLOE--MAMI discrepancy, we postulate the existence of the following \emph{nucleon–rescaled} low-energy operator that activates only in baryonic environments: 

\begin{equation}
  \label{eq:nucleon_vertex}
  \mathcal L_{\text{eff}}
   = g_{\text{eff}}\,
     (\bar N N)\,
     \varepsilon^{\mu\nu\alpha\beta}
     \bigl(\partial_\mu \mathcal{V}_{\!\mathcal B,\nu}\bigr)
     \bigl(\partial_\alpha \mathcal{A}_\beta\bigr)\,P ,
\end{equation}
where $(\bar N N)=\bar p\,p+\bar n\,n$ is the isoscalar nucleon density, $\mathcal V_{\!\mathcal B,\mu}$ is the new leptophobic vector potential, $\mathcal{A}_\beta$ the photon, $P\in\{\pi^0,\eta,\eta'\}$ a neutral pseudoscalar, and $\varepsilon^{\mu\nu\alpha\beta}$ the Levi-Civita tensor.

In terms of field strengths,
$V_{\!\mathcal B}^{\mu\nu}\equiv\partial^\mu \mathcal V_{\!\mathcal B}^{\nu}-\partial^\nu \mathcal V_{\!\mathcal B}^{\mu}$ and
$\tilde F_{\mu\nu}\equiv\tfrac12\varepsilon_{\mu\nu\alpha\beta} F^{\alpha\beta}$, Eq.~\eqref{eq:nucleon_vertex} is equivalently
\begin{equation}
  \label{eq:nucleon_vertex_compact}
  \mathcal L_{\text{eff}}
  \;=\;
  \frac{g_{\text{eff}}}{4}\,
  (\bar N N)\,V_{\!\mathcal B}^{\mu\nu}\,\tilde F_{\mu\nu}\,P.
\end{equation}
This operator switches on only in nucleon environments, vanishing identically in purely leptonic production channels.

Regarding the large-$N_{c}$ scaling, the Standard Model $VP\gamma$ vertex is $O(N_c^{0})$ as two mesons are present~\cite{Lebed1999}. Meson-baryon-antibaryon vertices scale individually as $O(\sqrt{N_c})$. However, the full meson-baryon scattering amplitude is constrained by the large-$N_{c}$ consistency relations to remain at most $O(N_c^{0})$, due to the cancellation of the leading $O(N_c)$ contributions between the $s$- and $u$-channel diagrams~\cite{Witten1979,Jenkins2009,Dashen1993ab,Jenkins1993abc,Dashen1994}. Therefore, the contribution arising from Eqn.~\eqref{eq:nucleon_vertex_compact} scales as $O(N_c^{0})$ and is \emph{not} further $1/N_c$-suppressed relative to the dominant $O(N_c^{0})$ Standard Model VMD term~\cite{Witten1979,Jenkins2009,Dashen1993ab,Jenkins1993abc,Dashen1994}.

\paragraph{UV completion.} 

This effective interaction can arise from ultraviolet (UV) sectors that have been widely discussed in the literature and are further elaborated below.

In addition to the above-mentioned leptophobic vector $V_{\!\mathcal B}$, we introduce a CP-even scalaron $S$ with a Yukawa coupling to nucleons:

\begin{equation}
    \mathcal{L_{\text{scalar}}} = g_S\,S\bar{N}N,
\end{equation}
where we assume $m_S \!\sim\! 1~\text{GeV}$. In this mass range, the scalar-nucleon Yukawa coupling $g_S$ is only weakly constrained, and can naturally take values large enough to address the KLOE–MAMI discrepancy.

Constraints from nanometer- and femtometer-scale fifth-force and nuclear-interaction searches lose sensitivity at sub-femtometer ranges~\cite{kamiya2015,xu2013}, while the supernova SN~1987A limits are strongly suppressed for $m_S \gtrsim 200~\mathrm{MeV}$~\cite{dev2020,hardy2025}. Consequently, both classes of searches do not constrain the $\rm{GeV}$-scale scalaron. In principle, an additional Yukawa component of the form $\frac{g_S e^{-m_S r}}{r}$ could be fitted directly to nuclear data~\cite{epelbaum2009}, but to the best of the author's knowledge, such a dedicated analysis has not yet been performed.

The existing collider limits on $\rm{GeV}$-scale CP-even scalars are model-dependent, typically assuming either Higgs-portal mixing~\cite{winkler2019,boiarska2019} or explicit couplings to quarks and gluons that generate nucleon-level interactions through QCD matching~\cite{batell2019,kling2023}. These studies probe partonic processes rather than interactions with nucleons as composite bound states.

In the absence of such partonic couplings, a purely \textit{nucleophilic} scalar with $m_S \sim 1~\mathrm{GeV}$ remains only mildly constrained, and can evade existing laboratory, astrophysical, and beam-dump  constraints~\cite{Knapen:2017,batell2019}. This leaves room for a comparatively large Yukawa coupling $g_S$, potentially strong enough to account for the observed KLOE–MAMI discrepancy.

The third component of our framework is a CP-odd dark pion, $\pi_D$, whose \rm{GeV}-scale phenomenology has been discussed in detail in~\cite{cheng2022}. This particle is motivated by the idea of a light, confining hidden sector that couples feebly to the Standard Model, an avenue broadly known as a hidden valley (HV)~\cite{strassler2007}. The previously introduced leptophobic vector $V_{\!\mathcal{B}}$ can originate from the same confining sector as a composite vector resonance, analogous to the QCD $\omega(782)$~\cite{harada2003}.

With an appropriate assignment of dark-quark charges, one obtains a mixed dark anomaly term of the form $\pi_D\,V_{\!\mathcal B}^{\mu\nu}\,\tilde F_{\mu\nu}$, which links the dark pion simultaneously to the visible and hidden vector fields, while making the purely visible anomaly term $\pi_D\,F^{\mu\nu}\,\tilde F_{\mu\nu}$ zero. Additionally, recent collider searches for long-lived dark pions, such as the MoEDAL-MAPP~\cite{Arifeen:2025DarkPion}, provide further motivation to explore this framework.

Finally, to connect these components into the effective vertex in Eqn.~\eqref{eq:nucleon_vertex}, we postulate the existence of a trilinear bridge term, $S\,P\,\pi_D$, that couples the scalaron $S$ simultaneously to the Standard Model pseudoscalar $P$ and to the dark pion $\pi_D$. This interaction is similar to the SM $f_0(500)\,\pi\,\pi$ coupling discussed in the previous Section, except that here the scalaron mediates between the SM and dark pseudoscalar sectors rather than between two visible pions. 

The resulting ultraviolet Lagrangian, incorporating all relevant components, reads:

\begin{align}
\mathcal L_{UV} \supset\;
& \underbrace{g_S\, S\,\bar N N}_{\text{nucleon portal}}
\;+\;
\underbrace{c_{SP}\, S\,P\,\pi_D}_{\text{trilinear bridge}}
\;+\;
\underbrace{\frac{C_{V_\mathcal{B}\gamma}}{f_{\pi_D}}\,\pi_D\,V_{\!\mathcal B}^{\mu\nu}\,\tilde F_{\mu\nu}}_{\text{mixed dark anomaly}}
\;
\label{UV_complete},
\end{align}
where $c_{SP}$ has mass dimension $1$, $C_{V_\mathcal{B}\gamma}$ is an anomaly coefficient fixed by dark-quark charges, and $f_{\pi_D}$ is the dark pion decay constant.

Integrating out $S$ and $\pi_D$ at tree level yields the nucleon-triggered contact operator:

\begin{equation}
\mathcal L_{\rm eff}\;\supset\;
\frac{g_S\,c_{SP}\,C_{V_\mathcal{B}\gamma}\;}{m_S^2\,m_{\pi_D}^{2}\, f_{\pi_D}} \,
(\bar N N)\,P\,V_{\!\mathcal B}^{\mu\nu}\tilde F_{\mu\nu},
\end{equation}
resulting in the effective coefficient in Eqn.~\eqref{eq:nucleon_vertex}:
\begin{equation}
    g_{\rm eff}\;=\;\frac{4g_S\,c_{SP}\,C_{V_\mathcal{B}\gamma}\;}{m_S^2\,m_{\pi_D}^2\,f_{\pi_D}}
    \label{Coupling_Const}
\end{equation}

Importantly, the Lagrangian in Eqn.~\eqref{UV_complete} permits:
\begin{equation}
    p\bar{p}\;\rightarrow\; S^{(*)}\;\rightarrow\; \pi_D + P\,,
\end{equation}
which proceeds non-resonantly in $p\bar p$ annihilation when $m_S<2m_p$. We also choose $m_S<m_{\pi_D}+m_P$ so that the intermediate $S$ cannot go on shell into $\pi_D\,+ P$. For the benchmark $m_S\simeq1~\mathrm{GeV}$, $m_{\pi_D}\gtrsim1~\mathrm{GeV}$, and $P\in\{\eta,\eta'\}$, this condition is automatically satisfied. If both $S$ and $\pi_D$ are defined as isoscalars, the transition $p\bar p \to \pi_D+\pi^0$ is isospin-violating and thus suppressed, while $p\bar p \to \pi_D+\eta^{(\prime)}$ is isospin-allowed.

If $\pi_D$ decays invisibly, the experimental signature would appear as:
\begin{equation}
    p\bar p \to P + \text{missing mass}\,
\end{equation}
Conversely, if $\pi_D$ decays visibly, it can be searched for by its corresponding visible final states~\cite{cheng2022}, for example, by PANDA~\cite{PANDA2021}.

\paragraph{Discrete symmetry $\mathbb Z_3$.}

We additionally impose a discrete $\mathbb Z_3$ symmetry under which: 
\begin{equation*}
q(S)=0,\qquad q(P)=1,\qquad q(\pi_D)=2,\qquad q(V_{\!\mathcal B})=1,
\end{equation*}
while all other Standard Model fields, including $N$ and $\gamma$, are neutral.

This charge assignment preserves the interactions in Eqn.~\eqref{UV_complete}, since:

\begin{align}
q(S\bar N N) &= 0 + 0 + 0 \equiv 0~(\mathrm{mod}\,3), \nonumber\\[4pt]
q(SP\pi_D)   &= 0 + 1 + 2 \equiv 0~(\mathrm{mod}\,3), \nonumber\\[4pt]
q(\pi_D\,V_{\!\mathcal B}^{\mu\nu}\,\tilde F_{\mu\nu})
              &= 2 + 1 + 0 \equiv 0~(\mathrm{mod}\,3),
\label{eq:z3charges}
\end{align}

At the same time, this $\mathbb Z_3$ symmetry prohibits the $P\,V_{\!\mathcal B}^{\mu\nu}\,\tilde F_{\mu\nu}$ term making this mechanism active exclusively in nucleon-triggered environments:
\begin{align}
q\!\big(P\,V_{\!\mathcal B}^{\mu\nu}\,\tilde F_{\mu\nu}\big) 
   &= 1 + 1 + 0 \equiv 2 \not\equiv 0~(\mathrm{mod}\,3),
\end{align}
thereby leaving the KLOE measurement unaffected and producing no effect in photon-fusion reactions such as 
$\gamma\!+\!\gamma\!\to\!\pi^0\!+\!\pi^0$ or 
$\gamma\!+\!\gamma\!\to\!\pi^0\!+\!\eta$, 
where the nucleon current is absent.  

Additionally, the direct coupling to the nucleon current, as well as the kinetic mixing are prohibited: 
\begin{align}
q\!\big(\mathcal{V}_{\!\mathcal B}^{\mu}\,\bar N\gamma_\mu N\big) 
   &= 1 + 0 \equiv 1 \not\equiv 0~(\mathrm{mod}\,3), \nonumber\\[4pt]
q\!\big(F^{\mu\nu}V_{\!\mathcal B,\mu\nu}\big) 
   &= 0 + 1 \equiv 1 \not\equiv 0~(\mathrm{mod}\,3),
   \label{Prohibited_terms}
\end{align}

Note that $(\bar N N)\,V_{\!\mathcal B}^{\mu\nu}\,\tilde F_{\mu\nu}\,P$ should be understood as an effective operator arising from the $\pi_D$ exchange, when the $\pi_D$ degree of freedom is kept explicit, it is $1\,+0\,+1\,+2\times 2 \equiv 0~(\mathrm{mod}\,3)$.

In general, $\mathbb Z_3$ is a common building block in model building~\cite{Lee2011,Feruglio2008,Ferreira2025}. It can arise as a residual subgroup of a broken gauged $\mathrm{U}\left(1\right)$ and is commonly used to prohibit certain decays, shape fermion mass hierarchies, and constrain the Yukawa sector.

\paragraph{Properties and observational signatures of $V_{\!\mathcal B}$.}

For the dark vector $V_\mathcal{B}$ in Eqn.~\eqref{eq:nucleon_vertex}, we assume it to be moderately heavy and focus on the mass window: 
\begin{equation}
\label{eq:mass_window}
1.5~\textrm{GeV}\;\lesssim\;m_{V_{\!\mathcal B}}\;\lesssim\;5~\textrm{GeV},
\end{equation}
since dedicated LHC dijet searches lose sensitivity below $m_{jj}\!\simeq\!5~\text{GeV}$~\cite{dobrescu2013,Dobrescu:2024}, while existing photoproduction measurements leave this $\rm{GeV}$-scale region largely unexplored~\cite{fanelli2017,Othman2024}. In this band, the $V_{\!\mathcal{B}}$ coupling remains weakly constrained and can therefore be large.

The lower bound in Eqn.~\eqref{eq:mass_window} could, in principle, be slightly relaxed, but must satisfy 
$m_{V_{\!\mathcal{B}}} > m_{\eta'}$  so that $\eta^\prime\!\not\to\!V_{\!\mathcal B}\gamma$ is kinematically prohibited. Otherwise, $\eta^\prime$ produced in $\gamma\,+\, p\!\,\to\,\!\eta^\prime\,+\, p$ or $\pi^-\,+\,p\!\,\to\,\!\eta^\prime\,+\,n$ would receive large non-standard contributions to its decays.

It is important to note that the proposed dark vector $V_{\!\mathcal B}$ lies in a similar mass range as higher $\omega$ excitations, such as
$\omega(1450)$, $\omega(1650)$, and $\omega(2220)$. However, these excited $\omega$ states have much smaller coupling constants than the ground state $\omega(782)$~\cite{Mojica2017,Piotrowska2017,Gutsche2009}, and therefore their contribution to the radiative decays $\eta^{(\prime)}\!\to\eta(\pi^0)\gamma\gamma$ is negligible. In contrast, our central hypothesis is that the dark vector
$V_{\!\mathcal B}$, while occupying a similar mass range, generates
a much stronger amplitude due to its strong effective coupling in Eqn.~\eqref{Coupling_Const}, which receives contributions from the nucleophilic scalar and the dark pion sectors, Eqn.~\eqref{UV_complete}. Both of these sectors can possess large couplings without conflicting with existing experimental bounds, allowing $V_{\!\mathcal B}$ to cause a significant observed deviation between the KLOE and MAMI measurements.

Our proposed \(V_{\!\mathcal B}\) resembles the leptophobic dark photon, $\mathcal{B}$, which is widely discussed in the literature~\cite{lee1955,pais1973,rajpoot1989,foot1989,he1990,carone1995,bailey1995,carone19951,aranda1998,nelson1989,tulin2014,Escribano:2022,Balytskyi:2023,barger1996}. But it departs from that setup in crucial aspects:

\begin{enumerate}\setlength\itemsep{0.3em}
\item It is $\sim{\cal O}(\mathrm{GeV})$ rather than sub-\rm{GeV} scale, and therefore has no significant mixing with the $\omega$ meson~\cite{Kamada:2025mix}.

\item Its interaction is explicitly proportional to $(\bar N N)$, and therefore, this mechanism is ``activated'' only in nucleon environments and automatically evades quarkonium constraints~\cite{aubert2009,ablikim2008}.

\item Unlike $\mathcal{B}$, the above-mentioned $\mathbb Z_3$ symmetry forbids any direct coupling of $V_{\!\mathcal B}$ to the nucleon current, Eqn.~\eqref{Prohibited_terms}, allowing it to interact only through the intermediate $\pi_D$ and $S$.

\end{enumerate}

To emphasize this distinction, we denote our proposed new, nucleon-triggered vector by \(V_{\!\mathcal B}\), reserving the symbol \(\mathcal B\) for the conventional leptophobic dark photon. In what follows, we take $V_\mathcal{B}$, like $\mathcal{B}$, to be isoscalar so that it affects charge-exchange and photoproduction reactions in the same way.

\paragraph{Direct photoproduction of $V_\mathcal{B}$.} 

\begin{figure}[h!] 
\includegraphics[width=0.5\textwidth]{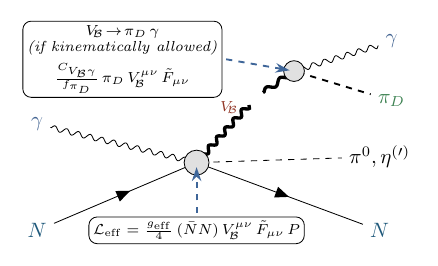}
\caption{Photoproduction of the dark vector $V_{\!\mathcal B}$ on a nucleon target,
$\gamma\,+\,N\!\to\!V_{\!\mathcal B}\,+\, P\,+\, N$, with the subsequent decay
$V_{\!\mathcal B}\!\to\!\pi_D\gamma$, if kinematically allowed.}
\label{Photoproduction_Diagram}
\end{figure}

It is important to note that while $\mathcal{B}$ is produced by diffractive $2\,\rightarrow\,2$ scattering photoproduction, as discussed in~\cite{fanelli2017,Othman2024}:
\begin{equation}
    \gamma\,+\,p \rightarrow \mathcal{B}\,+\,p,
\end{equation}
for $V_\mathcal{B}$ in Eqn.~\eqref{eq:nucleon_vertex}, it is a $2\,\rightarrow\,3$ process illustrated in Fig.~\ref{Photoproduction_Diagram}. For example, for photoproduction on the proton:
\begin{equation}
  \gamma + p \;\rightarrow\; V_{\!\mathcal B} + P + \gamma\,,\qquad P\in\{\pi^{0},\eta,\eta'\}\,
\end{equation}
If kinematically allowed, $m_{V_{\!\mathcal B}}>m_{\pi_D}$, the dark vector can subsequently decay as:
\begin{equation}
  V_{\!\mathcal B}\;\to\;\pi_D\,+\,\gamma\,, 
\end{equation}
via the mixed anomaly in Eqn.~\eqref{UV_complete}.

Depending on the dark pion parameters, exclusive $\gamma\,+\, p\!\to\!V_{\!\mathcal B}\,+\,P\,+\,p$ can manifest itself in three possible ways:
\begin{itemize}
  \item \textbf{Invisible:}  $P+p+\text{missing}$; if $m_{\pi_D}>m_{V_\mathcal{B}}$ and $V_{\!\mathcal B}$ is invisible.
  \item \textbf{Semi-visible:}  $P+p+\gamma+\text{missing}$; if $m_{V_\mathcal{B}}>m_{\pi_D}$ and $\pi_D$ is invisible.  
  \item \textbf{Fully visible:} $P+p+\gamma+(\pi_D\!\to\!X_{\rm SM})$, with $X_{\rm SM}\!\in\!\{\gamma\gamma,\ \ell^+\ell^-,\ \text{hadrons}\}$ as in ALP-like dark pion scenarios~\cite{cheng2022}, when $m_{V_\mathcal{B}}>m_{\pi_D}$.
\end{itemize}

For nuclear targets, the photoproduction cross section is expected to increase with the atomic number $\textrm{A}$, scaling as $\sigma_A \propto \rm{A}^2 $ in the coherent limit.

\begin{figure*}[!t]  
  \centering
  \includegraphics[width=\textwidth]{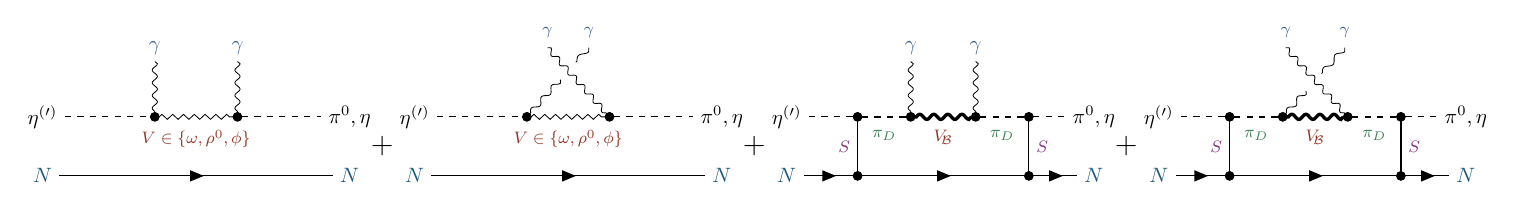} 
\caption{Dominant Standard Model and nucleon-triggered New Physics contributions to $\eta^{(\prime)}\!\rightarrow\!\pi^{0}(\eta)\gamma\gamma$ decays. In the Standard Model (left two diagrams), the nucleon acts as a spectator, and the decay proceeds via vector-meson dominance through intermediate $V\!\in\!\{\omega,\rho^{0},\phi\}$. In the New Physics sector (right two diagrams), the nucleon triggers the scalar portal $S$, which links the visible pseudoscalar $P$ to dark pions $\pi_{D}$ and to the isoscalar dark vector $V_{\!\mathcal B}$, generating an additional 
$V_{\!\mathcal B}P\gamma$ amplitude.}
\label{Fig_SM_and_VB_added}
\end{figure*}

\paragraph{Impact of $V_{\!\mathcal B}$ on $\eta^{(\prime)}\!\rightarrow\!\pi^0(\eta)\gamma\gamma$ decays.}

The $\eta$ and $\eta^\prime$ mesons are produced via $s$-channel nuclear resonance exchange and $t$-channel vector meson exchange~\cite{Benmerrouche1995,Chiang2003,Tiator2018}. Since the dark vector $\mathcal{V}_\mathcal{B}$ does not couple directly to the nucleon current, as shown in Eqn.~\eqref{Prohibited_terms}, and its mixing with the $\omega$ meson is negligibly small, we expect its impact on the production of $\eta$ and $\eta^\prime$ to be negligible.

The normalization channels $\eta\rightarrow\gamma\gamma$ and $\eta^\prime\rightarrow\gamma\gamma$ are unaffected by $V_\mathcal{B}$ due to its 
negligibly small mixing with the $\omega$ meson that participates in the axial anomaly~\cite{Feldmann1998}. Similarly, the normalization channels $\eta\!\to\!3\pi^0$ and $\eta^\prime\!\to\!3\pi^0$ are governed by the isospin-breaking component of the QCD Lagrangian rather than by vector boson dynamics~\cite{Gasser1985,Bijnens2007,Colangelo2018,Akdag2022,Borasoy2005}, and are therefore unaffected. In contrast, $V_{\!\mathcal B}$ can cause a significant modification in the doubly radiative decays $\eta^{(\prime)}\!\to\!\pi^0(\eta)\gamma\gamma$ in the presence of nucleon current, as discussed further in the text.

The new interaction in Eqn.~\eqref{eq:nucleon_vertex_compact} is effectively a ``nucleon-rescaled'' analogue of $\mathcal{B}$~\cite{tulin2014,fujiwara1985}. Unlike $\mathcal{B}$, however, $V_{\mathcal B}$ does not directly couple to the nucleon vector current, $\cancel{\mathcal{V}^\mu_\mathcal{B}\bar{N}\gamma_\mu N}$, as shown in Eqn.~\eqref{Prohibited_terms}, and the mixed anomaly in Eqn.~\eqref{UV_complete} is generated by the \emph{dark} quarks rather than visible ones. Since both $V_{\mathcal B}$ and $\mathcal{B}$ are isoscalar vectors, their flavor structure under the flavor $\mathrm{SU}\left(3\right)$ is identical, and thus the visible $V_{\mathcal B}P\gamma$ couplings can be parametrized in complete analogy with those of $\mathcal{B}$. Following~\cite{tulin2014,fujiwara1985}, the electromagnetic factor $e\,\text{Tr}\,[\mathbf Q\,\mathbf T_V]$ is replaced by:

\begin{equation}
  e\,\text{Tr}\bigl[\mathbf Q\,\mathbf T_V\bigr]
  \;\longrightarrow\;
  \frac{g_{V_\mathcal{B}}}{3}\,\text{Tr}\bigl[\mathbf T_V\bigr],
\end{equation}
where \({\bf Q}=\mathrm{diag}(\tfrac23,-\tfrac13,-\tfrac13)\) and
\({\bf T}_{V}\) is the \(\rm{U}(3)\) generator of the neutral vector. 

In the exact $\rm{SU}(3)$ and OZI limits, one finds:

\begin{equation}
  \frac{g_{V_\mathcal{B} P\gamma}}{g_{\omega P\gamma}}
  \;=\;
  \frac{(\tfrac{g_{V_\mathcal{B}}}{3})\,\mathrm{Tr}[T_\omega]}
       {e\,\mathrm{Tr}[T_\omega Q]}
  \;=\;
  2\,\frac{g_{V_\mathcal{B}}}{e},
\end{equation}
for \(P=\pi^{0},\eta,\eta^\prime\), which we compute using Eqn.~\eqref{CouplingConstants}.

In the isospin limit, we parameterize the squared matrix element corresponding to the vertex in Eqn.~\eqref{eq:nucleon_vertex} as:
\begin{equation}
   \bigl\langle p(n)\bigl|(\bar N N)^{2}\bigr|p(n)\bigr\rangle
      = C_{N}\,\bigl(2m_{N}\bigr)^{6},
\end{equation}
where the dimensionless constant \(C_{N}\) absorbs all possible nuclear corrections.

With two insertions of the nucleon portal vertex in Eqn.~\eqref{eq:nucleon_vertex}, the resulting contribution of the dark vector $V_{\!\mathcal B}$ to the $\eta^{(\prime)}\!\to\!\pi^{0}(\eta)\gamma\gamma$ amplitude reads:

\begin{equation}
\label{MVBAmplitude}
\begin{split}
\mathcal{M}_{V_{\mathcal{B}}}
&= \underbrace{\frac{g_{eff}^2\,C_N}{4\pi\,M^2_{V_{\mathcal{B}}}}}_{C_{\text{eff}}}
  \;\cdot\;\frac{4\,(2m_N)^6\,g_{\omega\eta^{(\prime)}\gamma}\,g_{\omega\pi^0(\eta)\gamma}}{\alpha_{\rm em}}\\
&\quad\times
  \Bigl[\bigl(P\!\cdot\!(q_1+q_2)-m_{\eta^{(\prime)}}^2\bigr)\{a\}
        \;-\;2\{b\}\Bigr],
\end{split}
\end{equation}
where the Lorentz structures \(\{a\}\) and \(\{b\}\) are defined in the same way as in Eqn.~\eqref{VMD}. The $\frac{1}{4\pi}$ factor arises since $g_{eff} \propto C_{V_\mathcal{B}\gamma}\propto g_{V_\mathcal{B}}$, hence $g^2_{eff} \propto\alpha_{V_\mathcal{B}} = \frac{g_{V_\mathcal{B}}^2}{4\pi}$, to compensate $4\pi$ from $\alpha_{\rm em}$.

Since the relevant momentum transfers satisfy \(m_{V_{\!\mathcal B}}^{2}\gg t,u\) in the mass range of Eqn.~\eqref{eq:mass_window}, we approximate the Breit–Wigner propagator in Eqn~\eqref{MVBAmplitude} as its leading constant term, \(1/m_{V_{\!\mathcal B}}^{2}\), and neglect the width \(\Gamma_{V_{\!\mathcal B}}\). In this approximation, the heavy-vector exchange acts as a local contact interaction, and the entire effect on all three decays $\eta^{\left(\prime\right)}\rightarrow\eta\left(\pi^0\right)\gamma\gamma$ is governed by a single parameter  \(C_{\text{eff}}\):
\begin{equation}
    C_{\text{eff}} = \frac{g_{eff}^2\,C_N}{4\pi\,M^2_{V_{\mathcal{B}}}},
\end{equation}
which we determine from the MAMI data in the next Section~\ref{Numerical}.

The amplitude $\mathcal{M}_{V_{\!\mathcal B}}$ 
has the same Lorentz structure as the Standard Model VMD term and interferes coherently with it. As shown in Fig.~\ref{Fig_SM_and_VB_added}, for the Standard Model vectors, the nucleon is a spectator, while the $V_{\!\mathcal B}$ exchange is activated by the nucleon through the intermediate scalar $S$ and the dark pion $\pi_D$. In addition to the VMD contribution shown in 
Fig.~\ref{Fig_SM_and_VB_added}, there are subleading pion and kaon loops, previously discussed, in which the nucleon remains a spectator as well.

This coefficient is universal for all three decays $\eta^{\left(\prime\right)}\rightarrow\pi^0\left(\eta\right)\gamma\gamma$. Once fixed by fitting the MAMI \(\eta\!\to\!\pi^{0}\gamma\gamma\) data~\cite{nefkens2014new}, we can make parameter–free predictions for \(\eta'\!\to\!\pi^{0}\gamma\gamma\) and \(\eta'\!\to\!\eta\gamma\gamma\) in the nucleon production modes (photoproduction or charge–exchange), where the nucleon trigger is present. These predictions differ from the Standard Model decay rates observed by BESIII~\cite{ablikim2017observation,ablikim2019search},
which correspond to purely leptonic production and hence do not activate the operator in Eqn.~\eqref{eq:nucleon_vertex}.

The fitting results and the associated falsifiable predictions are presented in the next Section~\ref{Numerical}.

\onecolumngrid\
\begin{table*}[htb]
  \centering
  \setlength{\tabcolsep}{6pt}
  \renewcommand{\arraystretch}{1.25}
  \begin{tabular}{|c|l|c|c|l|}
    \hline
    \textbf{Decay mode} & \textbf{Scenario} 
      & \(\Gamma_{\text{th}}\,[\text{GeV}]\) 
      & \(\text{BR}_{\text{th}}\) 
      & \(\text{BR}_{\text{exp}}\) \\ 
    \hline\hline

\multirow{3}{*}{$\eta\!\to\!\pi^{0}\gamma\gamma$}
  & SM, leptonic production 
  & $1.59(7)\times10^{-10}$ 
  & $1.22(7)\times10^{-4}$ 
  & $0.98(18)\times10^{-4}$ (KLOE \cite{Babusci2025}) \\ \cline{2-5}
  & SM + $V_{\!\mathcal B}$, nucleon production
  & $3.44(2)\times10^{-10}$ 
  & $2.65(10)\times10^{-4}$ 
  & $2.52(25)\times10^{-4}$ (MAMI\cite{nefkens2014new}), \\
  & 
  & 
  & 
  & $2.55(22)\times10^{-4}$ (PDG\cite{PDG}) \\
\hline

    \multirow{2}{*}{$\eta'\!\to\!\pi^{0}\gamma\gamma$}
      & SM, leptonic production
      & $5.41(46)\times10^{-7}$ 
      & $2.84(27)\times10^{-3}$ 
      & $3.20(24)\times10^{-3}$ (BESIII \cite{ablikim2017observation}) \\ \cline{2-5}
      & SM + \(V_{\!\mathcal B}\), nucleon production
      & $6.00(52)\times10^{-7}$ 
      & $3.15(31)\times10^{-3}$ 
      & --- \\ 
    \hline

\multirow{3}{*}{$\eta'\!\to\!\eta\gamma\gamma$}
  & SM, leptonic production 
  & $1.62(17)\times10^{-8}$ 
  & $0.86(10)\times10^{-4}$ 
  & $\mathrm{BR}<1.33\times10^{-4}$ (90\% C.L., BESIII \cite{ablikim2019search}), \\
  & 
  & 
  & 
  & $8.25(3.49)\times10^{-5}$ (2.6\,$\sigma$, BESIII \cite{ablikim2019search}) \\ \cline{2-5}
  & SM + $V_{\!\mathcal B}$, nucleon production
  & $1.59(17)\times10^{-8}$ 
  & $0.84(10)\times10^{-4}$ 
  & --- \\ 
\hline

  \end{tabular}

  \caption{Branching ratios and partial widths in the Standard Model
          and after including the nucleon–rescaled vector
           \(V_{\!\mathcal B}\).  The single fit to the MAMI
           \(\eta\!\to\!\pi^{0}\gamma\gamma\) data fixes the effective
           coupling \(C_{\text{eff}} = 9.47(53)\times10^{-5}\,\text{GeV}^{-8}\),
           which is then used to predict the corresponding rates in
           nucleon production environments for
           \(\eta'\!\to\!\pi^{0}\gamma\gamma\)
           and \(\eta'\!\to\!\eta\gamma\gamma\).}
  \label{tab:fitSummary}
\end{table*}

\onecolumngrid\
\begin{figure}[!t]
  \centering
  \setlength{\tabcolsep}{4pt}
  \begin{tabular}{cc}
    \includegraphics[width=0.50\textwidth]{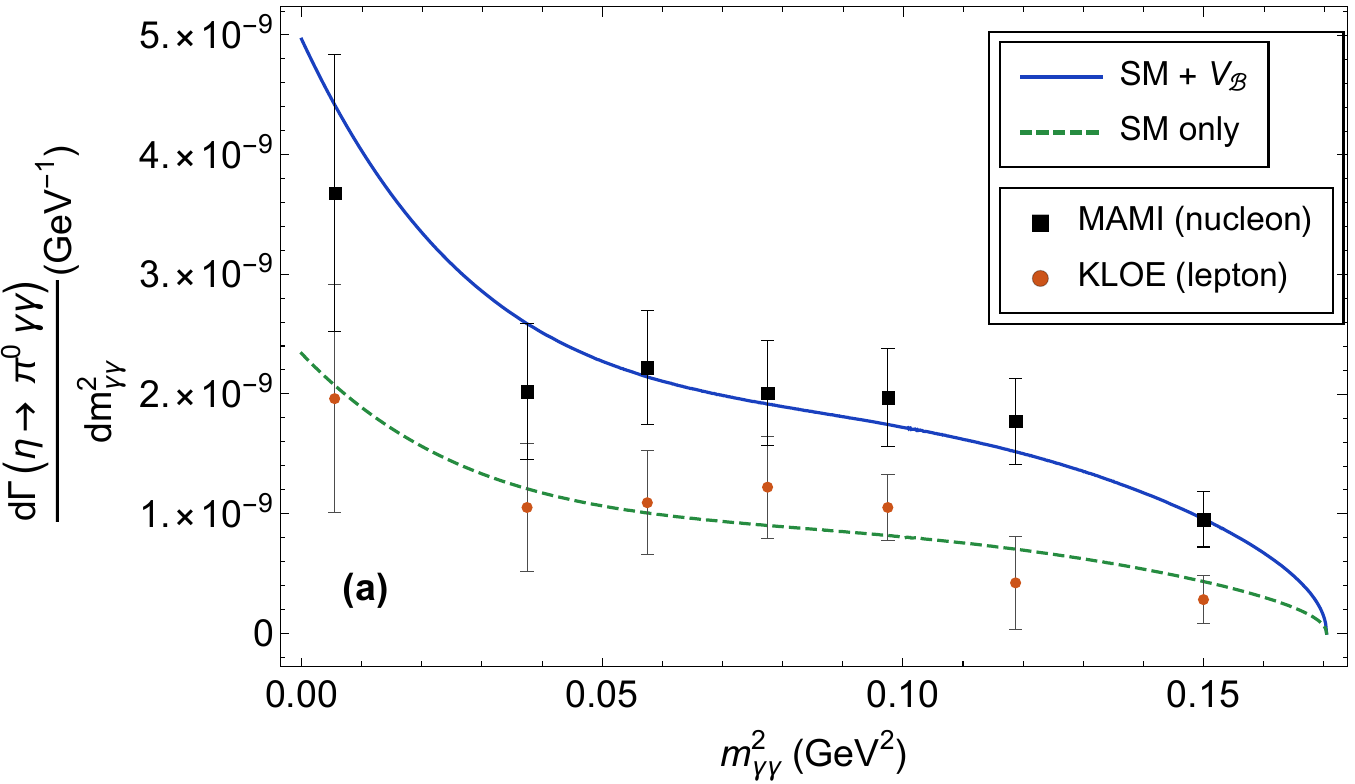} &
    \includegraphics[width=0.50\textwidth]{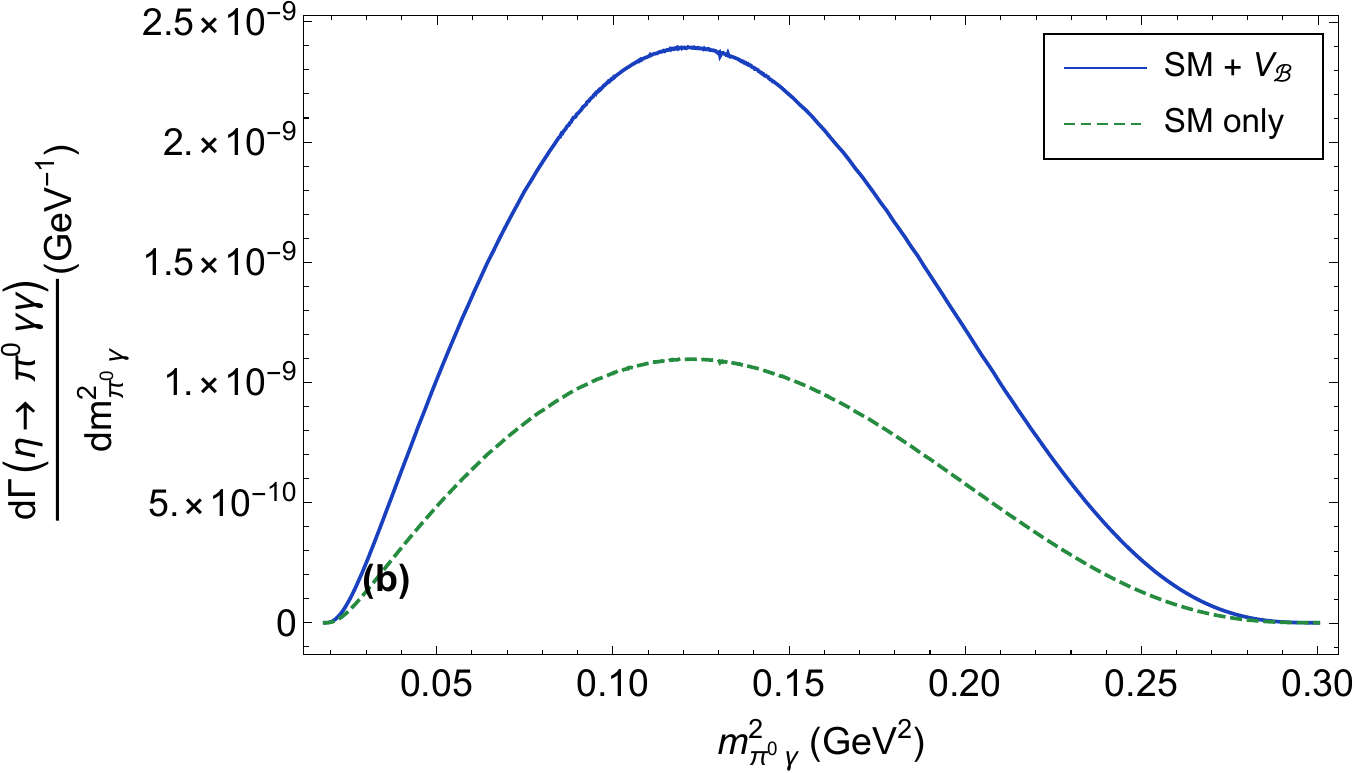}
  \end{tabular}
  \caption{Differential $\gamma\gamma$ spectrum in subfigure \textbf{(a)} and $\pi^{0}\gamma$ spectrum in subfigure \textbf{(b)} for the $\eta\!\to\!\pi^{0}\gamma\gamma$ decay. In both panels, the dashed green curve shows the Standard Model prediction, while the solid blue curve represents the 
nucleon-triggered scenario including $V_{\!\mathcal B}$. The new interaction induces an almost uniform upward shift across the spectra.}
  \label{EtaPiFitPic}
\end{figure}

\begin{figure}[!t]
  \centering
  \setlength{\tabcolsep}{4pt}
  \begin{tabular}{cc}
    \includegraphics[width=0.50\textwidth]{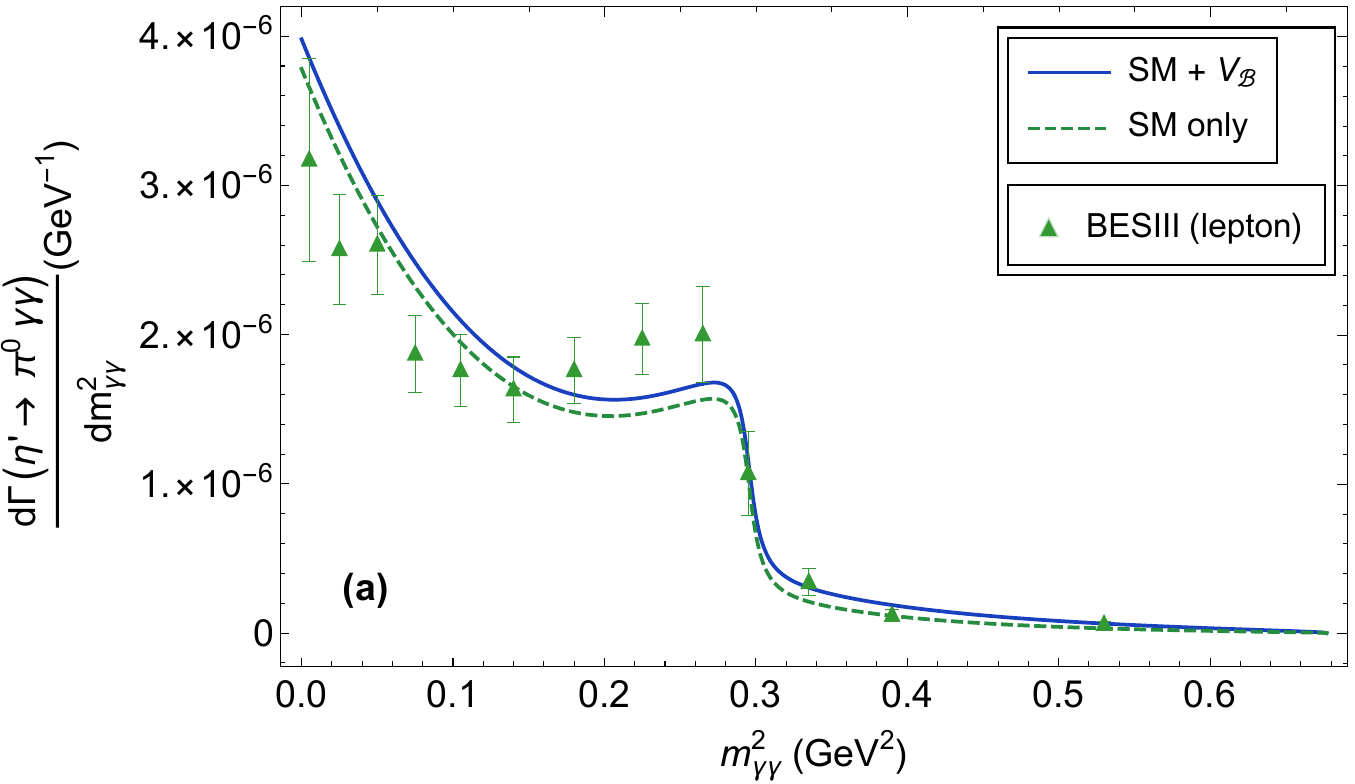} &
    \includegraphics[width=0.50\textwidth]{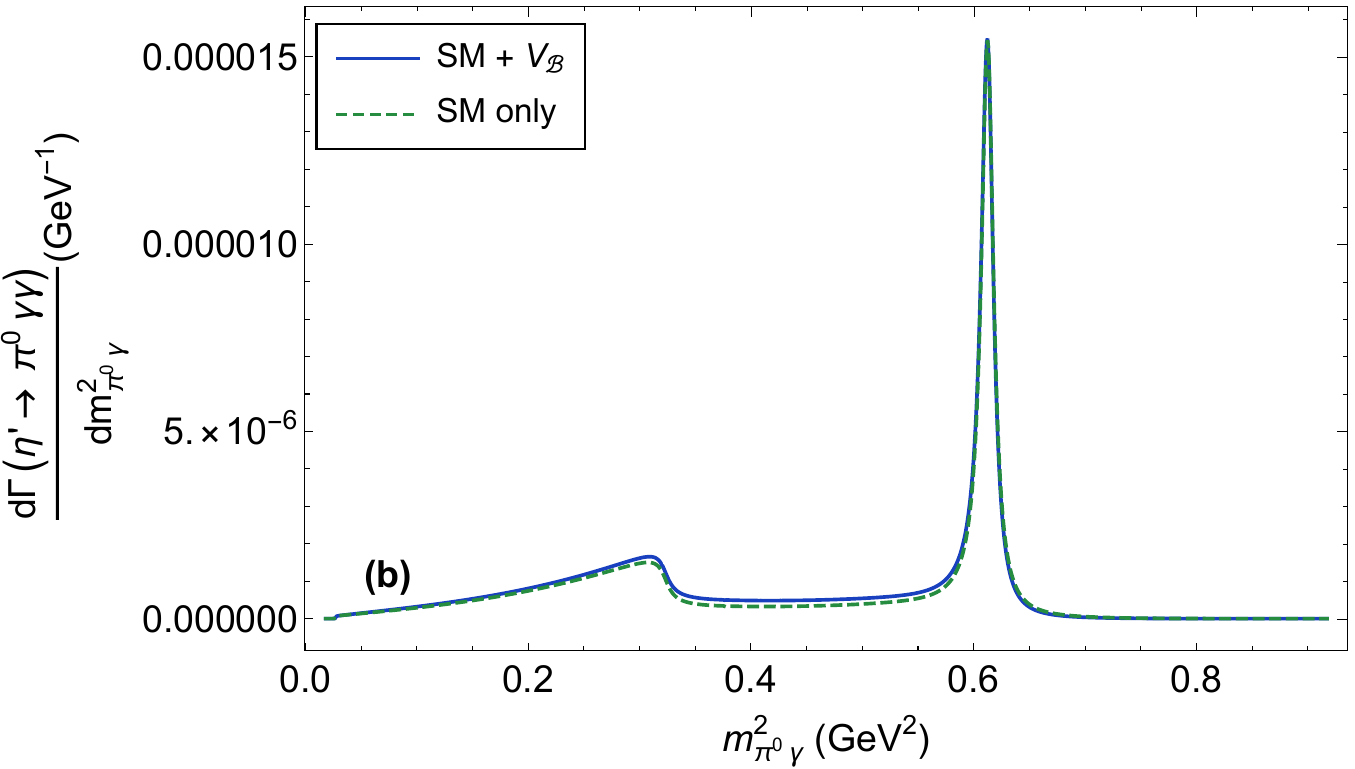}
  \end{tabular}
  \caption{Differential $\gamma\gamma$ spectrum in subfigure \textbf{(a)} and $\pi^{0}\gamma$ spectrum in subfigure \textbf{(b)} for the $\eta^\prime\!\to\!\pi^{0}\gamma\gamma$ decay. The $V_\mathcal{B}$ term produces a $\sim 10\%$ increase in the nucleon-triggered production environment (blue versus green), smaller than in the $\eta$ case, since the $\omega$ meson is on shell and dominates the overall Standard Model amplitude.}
  \label{EtaPrimePiFitPic}
\end{figure}

\begin{figure}[!t]
  \centering
  \setlength{\tabcolsep}{4pt}
  \begin{tabular}{cc}
    \includegraphics[width=0.50\textwidth]{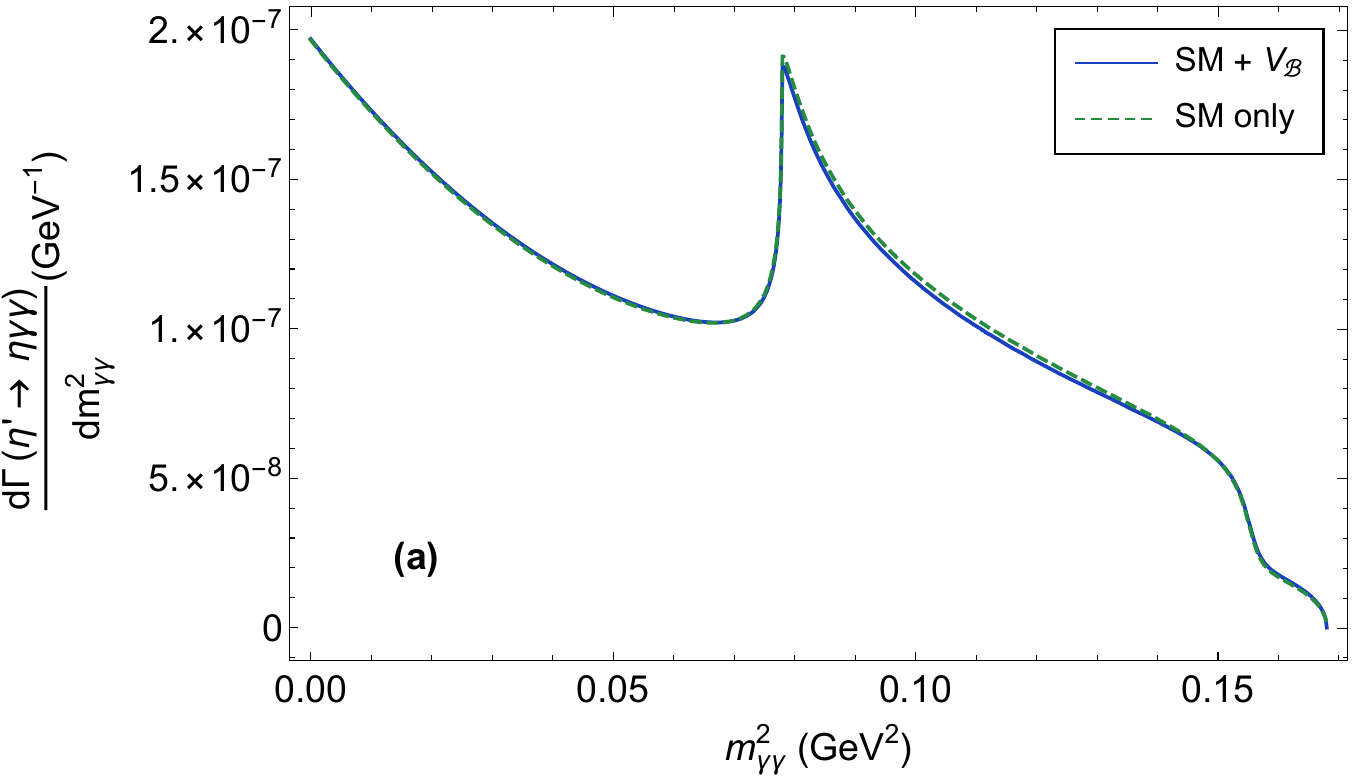} &
    \includegraphics[width=0.50\textwidth]{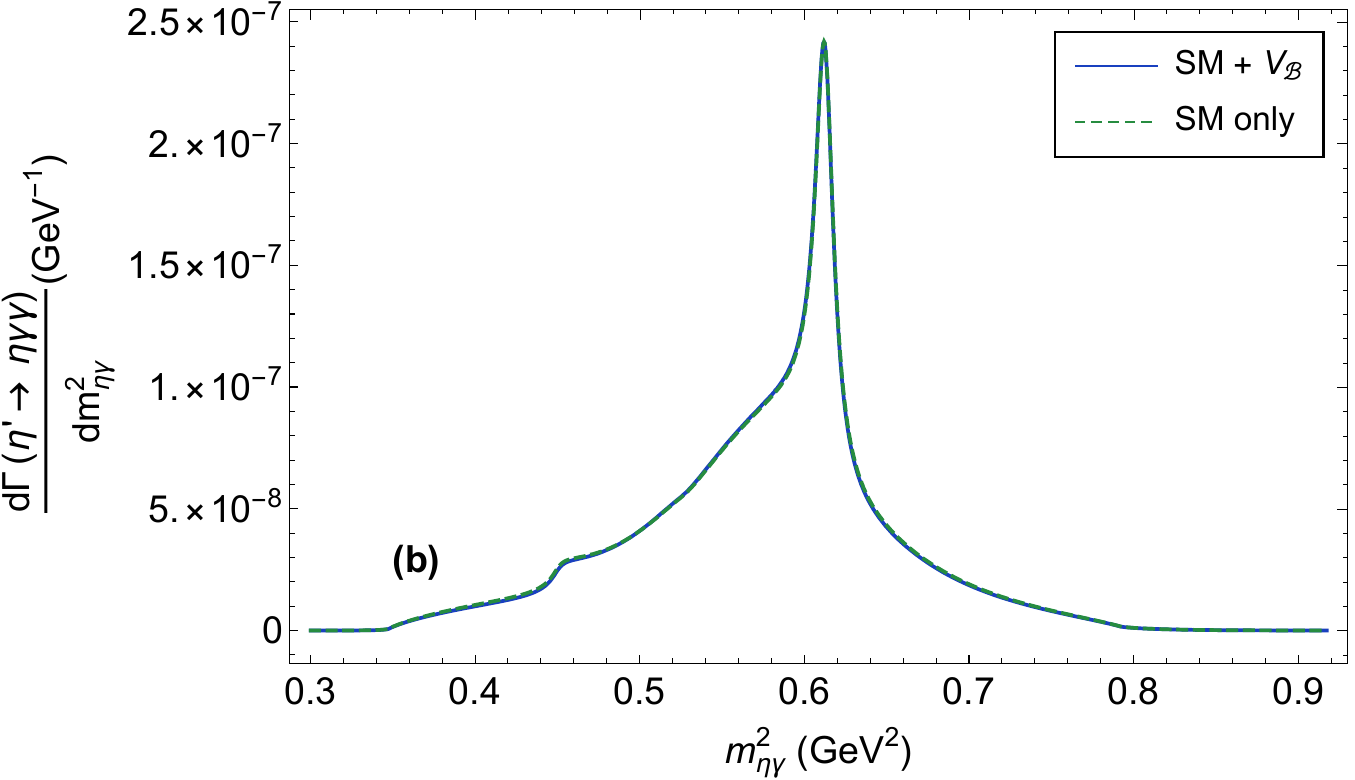}
  \end{tabular}
    \caption{Differential $\gamma\gamma$ spectrum in subfigure \textbf{(a)} and $\eta\gamma$ spectrum in subfigure \textbf{(b)} for the $\eta^\prime\!\to\!\eta\gamma\gamma$ decay. This decay is dominated by $\rho^{0}$, and in the nucleon production environment, the off-shell \(V_{\!\mathcal B}\) effect (blue) is at the $\sim1\%$ level, and negative.}
  \label{EtaPrimeEtaFitPic}
\end{figure}
\twocolumngrid\

\section{Numerical results}\label{Numerical}

We determine a single free parameter in our model \(C_{\text{eff}}\) by performing the fit to the MAMI data by minimizing:

\begin{equation}
\chi^2 = \sum_k \frac{\left(\frac{\mathrm{d}\Gamma_{\text{theory}}^{\eta\rightarrow\pi^0\gamma\gamma}\left[{{\cal M}^{\mathrm{VMD}} + \mathcal{M}^{L\sigma M}
+ \mathcal{M}_{V_{\mathcal{B}}}\left(C_{\text{eff}}\right)}\right]}{\mathrm{d}m^2_{\gamma\gamma}} - \frac{\mathrm{d}\Gamma_{\text{MAMI}}^{\eta\rightarrow\pi^0\gamma\gamma}}{\mathrm{d}m^2_{\gamma\gamma}}  \right)_k^2}{\sigma_k^2}
\end{equation}

The corresponding best-fit value is: 
\begin{equation}
C_{\text{eff}} = 9.47\left(53\right)  \times 10^{-5}\, \mathrm{GeV}^{-8},
\end{equation}
and the resulting spectra for $\eta\rightarrow\pi^0\gamma\gamma$ are shown in Fig.~\ref{EtaPiFitPic}.  Since \(V_{\!\mathcal B}\) is far off shell, in nucleon production environment, its contribution adds an almost constant offset to the Standard Model lineshape.

Using this resulting fit value, we predict the decay rates of $\eta^\prime\rightarrow\pi^0\gamma\gamma$ and $\eta^\prime\rightarrow\eta\gamma\gamma$ when measured in nucleon production environments, which has not yet been done at the time of writing this article. The corresponding numerical values are summarized in Table~\rm{II}.

For the $\eta^\prime\!\to\!\pi^{0}\gamma\gamma$ decay, the \(V_{\!\mathcal B}\) term increases the rate by about \(10\%\) in a nucleon environment (Fig.~\ref{EtaPrimePiFitPic}).  The relative effect is smaller than for $\eta\!\to\!\pi^{0}\gamma\gamma$ since, in the $\eta'$ decay, an on-shell $\omega$ can resonate on the Dalitz plot giving a strong contribution that dominates the Standard Model amplitude, whereas $V_{\!\mathcal B}$ is far off shell and is therefore suppressed. On the other hand, in the $\eta\rightarrow\pi^0\gamma\gamma$ decay, both the Standard Model particles, $\omega$ and $\rho^{0}$, and the new vector are off-shell, and therefore, their relative impacts are comparable.

For the case of the $\eta'\!\to\!\eta\gamma\gamma$ decay, the $V_{\!\mathcal B}$ contribution is negligible, even in the nucleon production environment, shifting the decay width \textit{downwards} by only $\sim\!-1\%$ (Fig.~\ref{EtaPrimeEtaFitPic}). This channel is dominated by the $\rho^{0}$ meson rather than the $\omega$, since numerically:  
\begin{equation}
   g_{\omega\eta\gamma}g_{\omega\eta^\prime\gamma} \ll g_{\rho^0\eta\gamma}g_{\rho^0\eta^\prime\gamma},
\end{equation}
and the same hierarchy suppresses the $V_{\!\mathcal B}$ term in Eqn.~\eqref{MVBAmplitude}. Among the three decays considered, $\eta^{\prime}\!\to\!\eta\gamma\gamma$ is therefore the least sensitive to this type of New Physics scenarios.

Finally, the Standard Model spectrum in Fig.~5 deviates slightly from that provided in~\cite{escribano2020theoretical}. The difference stems from our updated parametrization of the $\rho^{0}$ line shape, which is particularly relevant here since $\rho^{0}$ provides the dominant contribution to this decay.

In the next Section~\ref{Discussion}, we outline additional observables that can confirm or refute this nucleon-rescaled vector scenario.

\section{Discussion}
\label{Discussion}

If the proposed nucleon-triggered $V_{\!\mathcal B}$ mechanism is correct, the upcoming
Jefferson Lab Eta Factory (JEF)~\cite{JEFproposal}, which also produces $\eta$ mesons via $\gamma p\to\eta p$, is expected to find a value consistent with the MAMI result~\cite{nefkens2014new}. In contrast, BESIII~\cite{BESIII2023JpsiRadiative} produces $\eta$-mesons through $e^+e^-\!\to\!J/\psi\!\to\!\gamma\eta$, without an external nucleon current, so our mechanism does not contribute there. Thus, if $\eta\rightarrow\pi^0\gamma\gamma$ is measured in the future using BESIII, our model predicts a value consistent with the KLOE measurement~\cite{Babusci2025}. A pattern in which photoproduction and charge-exchange experiments agree with each other but differ from leptonic-production experiments would provide a sharp test of our model and could confirm or exclude it.

For the mass window in Eqn.~\eqref{eq:mass_window}, we have
\(m_{V_{\!\mathcal B}}\!>\!m_{\eta},m_{\eta'},m_{\omega},m_{\phi}\), and unlike the case of sub-\rm{GeV} leptophobic dark photon $\mathcal{B}$, the radiative channels:
\begin{equation}
   \eta^{\left(\prime\right)} \not\rightarrow V_{\mathcal{B}}\gamma,
\end{equation}
are kinematically prohibited, and the $\omega-V_{\mathcal{B}}$ mixing is negligibly small. Therefore, the linear in $\rm{A}$ term in Eqn.~\eqref{eq:nucleon_vertex} has no significant effect on the two-particle decays of $\eta$, $\eta^\prime$, $\omega$, and $\phi$. The only phenomenologically relevant term has the scaling \(\mathrm{A}^{2}\), which adds nearly a constant offset to the decay rates whenever a nucleon current is present.

Since the New Physics amplitude scales as \(\langle\bar N N\rangle^{2}\sim
\rm{A}^{2}\), the signal should rise sharply with the mass number of the
target nucleus.  A decisive test is thus a systematic
study of:
\begin{equation}
   \gamma \rm{A}\;\longrightarrow\;(\eta,\eta')\,\rm{A}\;
       \longrightarrow\;\pi^{0}(\eta)\,\gamma\gamma\,\rm{A},
\end{equation}
which is possible with the current experimental setups~\cite{ElsaMAMI2012,Krusche2005,Shepherd2024,Paryev2012,Pheron2012}. Comparing the $\eta\rightarrow\pi^0\gamma\gamma$ decay rate for $\eta$ mesons photoproduced on light as opposed to heavy targets provides a direct test of the predicted \(\rm{A}^{2}\) enhancement, which would confirm or exclude the proposed model.

The model also predicts a \(\sim\!10\%\) upward shift in the
\(\eta'\!\to\!\pi^{0}\gamma\gamma\) branching ratio under nucleon
production (Table~\rm{II}). Testing this requires improved precision in both leptonic and nucleon production modes, since current experimental uncertainties are of comparable size. In contrast, the effect in \(\eta'\!\to\!\eta\gamma\gamma\) is opposite in sign to \(\eta^{\left(\prime\right)}\!\to\!\pi^{0}\gamma\gamma\) and much smaller, only \(\mathcal{O}(1)\%\). Detecting or ruling out such a deviation would require significantly better (ideally sub-percent) accuracy than is currently available.

GlueX has already studied $\gamma p\!\to\!\omega\,\pi^0 p$~\cite{Scheuer2024} and $\gamma p\!\to\!\omega\,\eta p$~\cite{BarrigaThesis}, which are kinematically and methodologically close to our proposed search $\gamma p\!\to\!V_{\!\mathcal B} P p$ with $P\in\{\pi^{0},\eta,\eta'\}$, using the same exclusive $2\!\to\!3$ reconstruction. In our case, the visible $\omega\!\to\!3\pi$ line shape is replaced by $V_{\!\mathcal B}$ decays that are invisible, semi-visible, or fully visible. A targeted scan for a narrow leptophobic vector in the mass window of Eqn.~\eqref{eq:mass_window} can discover or exclude this parameter space. A complementary search for a GeV-scale scalar coupled to nucleons and dark pion would probe the portal dynamics underlying Eqn.~\eqref{eq:nucleon_vertex}.

\section{Conclusions}\label{Conclusions}

\(\eta^{(\prime)}\!\to\!\pi^{0}(\eta)\gamma\gamma\)  decays provide a remarkably clean window onto possible leptophobic New Physics scenarios. An intriguing new measurement by KLOE, exhibiting a $5.5\,\sigma$ tension with the MAMI result, cannot be explained by conventional sub-\rm{GeV} leptophobic dark photon scenarios. Such models predict identical effects in both leptonic and nucleon-induced production environments, and are already tightly constrained by existing experimental data. In contrast, we propose a production-dependent, nucleon-rescaled vector boson \(V_{\!\mathcal B}\), whose mass lies in the \(m_{V_{\!\mathcal B}} = 1.5\text{--}5~\mathrm{GeV}\) range. This placement avoids the strict constraints that rule out conventional sub-\(\mathrm{GeV}\) dark photon scenarios. A \(\mathrm{GeV}\)-scale scalar and dark pion portal avoids the loop suppression of the \(V_{\!\mathcal B}\) coupling, providing the strong interaction needed to account for the sharp KLOE--MAMI discrepancy. Most importantly, this interaction is selectively activated in nucleon-induced production channels, leaving leptonic processes unaffected. A single fit to the MAMI \(\eta\!\to\!\pi^{0}\gamma\gamma\) spectrum determines a universal parameter that governs all related \(\eta^{(\prime)}\rightarrow\pi^0\left(\eta\right)\gamma\gamma\) decays, providing falsifiable predictions within reach of the current or near-future facilities. A dedicated direct photoproduction search for a narrow nucleon-philic vector in the \(1.5\text{--}5~\mathrm{GeV}\) mass range and its associated $\textrm{GeV}$-scale scalar and dark pion companions could either uncover the missing pieces of the model or decisively exclude the nucleon-triggered solution proposed here.

\section*{Code availability and reproducibility of our results}
\label{Reproducibility}

All our numerical code is publicly available on GitHub repository~\cite{NumericalCode}, to facilitate the use of our results.

\end{document}